\documentclass[%
 reprint,
 amsmath,amssymb,
 aps,
]{revtex4-2}

\usepackage{graphicx}% Include figure files
\usepackage{dcolumn}% Align table columns on decimal point
\usepackage{bm}% bold math
\usepackage{acro} % usage: \ac{ifo}
\DeclareAcronym{LQG}{short = LQG, long  =linear quadratic Gaussian}
\DeclareAcronym{DARM}{short=DARM, long=differential-arm length}
\DeclareAcronym{LIGO}{short=LIGO, long=Laser Interferometer Gravitational-Wave Observatory}
\DeclareAcronym{DOF}{short=DOF, long=degree of freedom, long-plural-form=degrees of freedom}
\DeclareAcronym{MIMO}{short=MIMO, long=multiple-input and multiple-output}
\DeclareAcronym{SISO}{short=SISO, long=single-input and single-output}
\DeclareAcronym{LSC}{short=LSC, long=LIGO Scientific Collaboration}
\DeclareAcronym{BBH}{short=BBH, long=binary black hole}
\DeclareAcronym{BNS}{short=BNS, long=binary neutron star}
\DeclareAcronym{FOM}{short=FOM, long=figure of merit, long-plural-form=figures of merit}
\DeclareAcronym{RMS}{short=RMS, long=root mean squared}
\DeclareAcronym{GW}{short = GW, long  = gravitational wave}
\DeclareAcronym{PSD}{short = PSD, long  = power spectral density}
\DeclareAcronym{ASC}{short = ASC, long  = alignment sensing control}
\DeclareAcronym{GPS}{short = GPS, long  = global positioning system}
\DeclareAcronym{SNR}{short = SNR, long  = signal-to-noise ratio}
\DeclareAcronym{BH}{short = BH, long  = Berstein-Haddad}
\DeclareAcronym{LMI}{short = LMI, long  = linear matrix inequality, long-plural-form=linear matrix inequalities}

\usepackage{autofigs12}
\graphicspath{{Graphics/}}
\usepackage{fontawesome}
\usepackage{amsmath}
\usepackage{bm}
\usepackage{braket}

\begin{document}

\newcommand{\fixme}[1]{{\textcolor{red}{[FIX: #1]}}}
\newcommand{\IM}[1]{{\color{blue}{\texttt{[IM: #1]}}}}
\newcommand{\LM}[1]{{\color{purple}{\texttt{[LM: #1]}}}}
\newcommand{\controller}{\ensuremath{K}}
\newcommand{\controllerSS}{\ensuremath{\mathbf{K}}}
\newcommand{\rd}{\ensuremath{\mathrm{d}}}

\newcommand{\rms}{\ensuremath{\tilde{R}}}

\newcommand{\trans}{\ensuremath{T}}

\newcommand{\approptoinn}[2]{\mathrel{\vcenter{
      \offinterlineskip\halign{\hfil$##$\cr
        #1\propto\cr\noalign{\kern2pt}#1\sim\cr\noalign{\kern-2pt}}}}}

\newcommand{\appropto}{\mathpalette\approptoinn\relax}

\preprint{APS/123-QED}

\title{Robust Bilinear-Noise-Optimal Control for Gravitational-Wave Detectors:\\A Mixed LQG/$\mathcal{H}_\infty$ Approach}% Force line breaks with \\

\author{Ian A. O. MacMillan}
 \email{macmillan@caltech.edu}
\author{Lee P. McCuller}%
\affiliation{%
 California Institute of Technology, PMA, 1200 E. California Blvd. Pasadena, CA 91125}%

\date{\today}% It is always \today, today,

\begin{abstract}
At its lowest frequencies, LIGO is limited by noise in its many degrees of freedom of suspended optics, which, in turn, introduce noise in the interferometer through their feedback control systems. Nonlinear interactions are a dominant source of low-frequency noise, mixing noise from multiple degrees of freedom. The lowest-order form is bilinear noise, in which the noise from two feedback-controlled subsystems multiplies to mask gravitational waves. Bilinear couplings require control trade-offs that simultaneously balance high- and low-frequency noise. Currently, there is no known lower limit to bilinear control noise. Here, we develop benchmark cost functions for bilinear noise and associated figures of merit. Linear-quadratic-Gaussian control then establishes aggressive feedback that saturates the lower bounds on the cost functions. We then develop a mixed LQG and $\mathcal{H}_\infty$ approach to directly compute stable, robust, and optimal feedback, using the LIGO's alignment control system as an example. Direct computations are fast while ensuring a global optimum. By calculating optimal robust control, it is possible to construct the lower bound on controls noise along the Pareto front of practical controllers for LIGO. This method can be used to drastically improve controls noise in existing observatories as well as to set subsystem control noise requirements for next-generation detectors with parameterized design.
% \begin{description}
% Secondary publications and information retrieval purposes.
% You may use the \texttt{description} environment to structure your abstract;
% use the optional argument of the \verb+\item+ command to give the category of each item. 
% \end{description}
\end{abstract}

\keywords{gravitational-wave detectors, LIGO, controls, bilinear noise, LQG optimization, mixed sensitivity, $\mathcal{H}_\infty$ optimization, robust control}%Use showkeys class option if keyword
\maketitle

%\tableofcontents
\section{Introduction}
\label{sec:intro}

The \ac{LIGO} is a ground-based gravitational wave interferometer that first detected gravitational waves in 2015~\cite{ligofirst}. Since its first detection, the \ac{LSC} has been working to reduce noise in the instrument to increase its sensitivity and, therefore, detect even fainter gravitational wave signals~\cite{O3sensitivity, O4SensitivityCapote}. These significant improvements in \ac{LIGO}'s sensitivity, as well as the inclusion of additional detectors, have led to an increased detection rate that expands astrophysical study~\cite{O3sensitivity, O4SensitivityCapote, O3aCatalouge, O3bCatalouge, O3aGRTests, O3bGRTests}.

There is considerable interest in reducing low-frequency noise in \ac{LIGO} ($<$30~Hz). There are many low-frequency waveforms, like high-mass \ac{BBH} inspirals, whose detection would benefit from an increased sensitivity in the low-frequency band~\cite{hall2021low-fq_sen}. Other waveforms that extend to higher frequencies will benefit from higher signal-to-noise detections, improved parameter estimation, and potentially early warning of ongoing events during the low-frequency portion of long-lived waveforms. Currently, at these low frequencies, both \ac{LIGO} detectors are limited by noise introduced by the feedback control loops. Improving the controllers used in \ac{LIGO}'s feedback loops and generally improving the processes for creating these controllers could significantly reduce the low-frequency noise in the detectors as well as allowing for resources to be spent on other areas of improvement. 

\subsection{Motivation}
The \ac{LIGO} instrumentation community widely uses classical control design methods. While an entire system has many degrees of freedom, they are almost all designed and engineered as individual \ac{SISO} signal paths using transfer functions, Bode plots, and frequency-resolved budgets of noise contributions. These subsystems and control loops are designed depending on how they interconnect to affect the principal measurement channels. The software interfaces of \ac{LIGO} are optimized for this approach, and it is largely successful; however, it has several weaknesses. First, the loops are all hand-designed, which can get the system operating fast, but compounds negatively for global optimization. In particular, without a lower bound on performance, a halting criterion for optimizing is established either by experience or by exhaustion. 
%Automated budget-optimization routines are sometimes employed, but are fragile to maintain and have few performance guarantees. 
Furthermore, the design of coupled \ac{MIMO} systems must often be iterative, as loop shapes and noise consequences are interdependent. More modern state-space controls approaches are an attractive research topic, promising guaranteed existence, stability, and optimality properties even for complicated, inter-coupled \ac{MIMO} systems. 

So far, modern controls approaches have not been widely adopted due to the unique requirements of gravitational-wave detectors. First, these approaches require very specific cost functions to form well-posed optimization problems. Second, these approaches usually assume white noise inputs; thus, system models must be augmented with noise-shaping filters. The fitting of such filters is nontrivial but now reliable with modern algorithms. The third issue is that interferometers rarely have actuator cost functions (unlike, e.g., aerospace); thus, typical examples of regulator cost do not immediately apply. The fourth issue is that interferometer loops are limited by transmission delays, which create nonminimum-phase zeros that must be incorporated into the control loop design. The final and highly significant hurdle is dynamic range. The range of shaped noise in gravitational-wave detectors spans 8-12 orders of magnitude, from low-frequency seismic motion to interferometric sensors. The combination of these factors causes optimal solutions to sit on the edge of stability regions, where even the best-known numerical algorithms struggle.

\subsection{Approach}
This manuscript adopts an intermediate and transitional approach in an effort to supplant hand-tuning while managing these challenges. It uses modern methods to pose and solve a \ac{SISO} controls problem. Future work can extend its examples to more intrinsically \ac{MIMO} problems. A major focus of this work is the problem definition, to pose the modern controls problem using \acp{FOM} that are relevant to optimize gravitational-wave detectors. Furthermore, the mixed-sensitivity approach we develop is necessary to find robust solvers that have acceptable gain and phase margins. This approach is mathematically complex to describe, but has largely been solved in previous controls literature. By finding solutions with acceptable phase margins, much like our hand-designed controllers, the numerical solutions are generally better behaved and do not live ``near the edge'' of admissible solution regions, where small perturbations, numerically or in the instrument plants, push them outside. Our mixed-sensitivity solver rapidly arrives at controllers qualitatively similar to hand designs but guaranteed optimal for our defined \acp{FOM} that are tied to the astrophysics objectives.

The organization of this paper is to be a bridge between classical and modern controls approaches. The first half provides the problem setup and results using the language of transfer functions and noise projections. Our results are included in these sections, showing how to form a modern controls optimization problem to minimize bilinear control noise and the resulting sets of controllers. The optimal controllers set bounds on the best possible noise performance, useful for determining a halting point for hand-designs. We then transform the problem description to use a mixed-sensitivity approach, yielding controllers with acceptable and adjustable stability margins. The second half of the paper translates the problem and solutions into state-space form and summarizes the literature and approach required to build our custom mixed-sensitivity numerical solver. This combination of approaches establishes a framework to utilize modern control systems to solve open controls challenges in gravitational-wave detectors.

\section{Background}
\label{sec:background}

\ac{LIGO}'s sensitivity is limited by the sum of a number of individual noise sources in the interferometer. The extent to which these noise sources obscure astrophysical signals is expressed using the \ac{SNR}. The \ac{SNR} of a signal can be calculated from the Fourier transform of the event waveform, $h_\mathrm{gw}(f)$, and the \ac{PSD} of the interferometer's noise, $S_h(f)$, using
\begin{align}\label{eq:detSNR}
   \mathrm{SNR}^2 &= 4\int_{0}^{\infty} \frac{\left |h_\mathrm{gw}(f)\right |^2}{S_h(f)} df \mathrm{.}
\end{align}
The detection range,
\begin{align}\label{eq:d_range}
  d_{\mathrm{range}} &= \frac{d_{\mathrm{gw}}}{2.26}\frac{\mathrm{SNR}}{\rho_0}\mathrm{,}
\end{align}
is the average effective distance to which events louder than a threshold \ac{SNR} called $\rho_0$ can be detected. The range can be calculated using the \ac{SNR} formula and the distance to the fiducial source. For an $h_\mathrm{gw}(f)$ waveform modeled from a source at distance $d_{\mathrm{gw}}$, the detection range averaged over the sky is included above. The threshold \ac{SNR} is usually taken to be $\rho_0 = 8$, and the factor of 2.26 accounts for the average of the interferometer antenna pattern over the sky~\cite{FinnPRD93ObservingBinary}.

The astronomical range is essential to the sensitivity of the interferometer because it dictates the detection rate, $\mathcal{R}_{\mathrm{det}}$, which holds the relations
\begin{align}
   \mathcal{R}_{\mathrm{det}} \propto d_{\mathrm{range}}^3 \propto \mathrm{SNR}^3 \mathrm{.}
\end{align}
Even minor decreases in the interferometer's noise can have a significant impact on the detection rate. One of \ac{LIGO}'s long-standing goals is increasing the detection rate for \ac{BNS} inspirals. Binary neutron star waveforms are taken to be the standard fiducial for computing \ac{SNR} as a \ac{FOM} for the interferometer performance in this work.

\subsection{Noise from Control Systems}
\label{subsec: controls noise}

At the low frequency end of its detection band, \ac{LIGO}'s sensitivity is limited by noise injected by control systems. At frequencies below \ac{LIGO}'s detection band, the control systems are stabilizing the operating points of otherwise highly nonlinear optical cavity resonators and interferometers. Noise coming from the control actuation of cavity alignment and auxiliary length are two of the largest of all known noise sources in the detector below 10~Hz~\cite{O3sensitivity, O4SensitivityCapote}. Interferometric controls noise is any noise that arises from the addition of a stabilizing or noise-suppressing controller to a \ac{DOF} of the interferometer system. When a controller is added in a feedback loop to a \ac{DOF}, some of the noise inherent to that \ac{DOF}, along with new measurement noise, passes through the controller and is (re-)injected into the system at the actuator. While a controller can reduce the overall noise of a \ac{DOF}, it also introduces a new noise source from the controller's sensor, particularly at higher frequencies corresponding to the feedback bandwidth. For complex systems with many \acp{DOF} and many control systems, the controls noise can become a significant noise source. From optics controllers to seismic isolation, each \ac{LIGO} detector has more than 300 controllers that all maintain its operating point but add noise to the principal readout: the \ac{DARM}. This is the measurement of arm-length differences caused by gravitational waves, and thus the output of interest for astrophysics.

Here, a model is constructed for how a control system for a single \ac{DOF} affects the signal and introduces noise. We will later express bilinear noise in terms of this model. The plant noise of some degree of freedom under feedback control, $N_{\mathrm{c}}$, affects the signal seen at the output of the interferometer through
\begin{align}
 h_\mathrm{signal}(f) &= h_\mathrm{gw}(f) + N_{\mathrm{det}}(f) + C(f) N_{\mathrm{c}}(f) \mathrm{,} 
 \label{eq: h_gw}
\end{align}
where $h_\mathrm{gw}(f)$ is a potential signal, $N_{\mathrm{det}}$ is all the detector noise from sources other than the \ac{DOF} being modeled, and $C(f)$ is a frequency-dependent coupling of the \ac{DOF} into \ac{DARM}. This coupling is often indirect and sourced from bilinear effects. For simple estimates, it is often possible to assume $C(f)$ is constant across all frequencies, but in the general case, measurement data necessitates that it be frequency-dependent. An example measurement of $C(f)$ and physical motivation are described in Section~\ref{subsec:foms}. The \ac{PSD} of the total noise in the detector is thus
\begin{align}
\label{eq:S_h}
 S_h(f) &= S_{\mathrm{det}}(f) + |C(f)|^2 S_{\mathrm{c}}(f) \mathrm{,}
\end{align}
where $S_{\mathrm{det}}(f)$ is the \ac{PSD} of all the detector noise from sources other than the \ac{DOF} being modeled and $S_{\mathrm{c}}(f)$ is the \ac{PSD} of the noise of interest, in this case a definition of controls noise. By linearizing Eq.~\ref{eq:detSNR}, and including Eq.~\ref{eq:S_h}'s expression of $S_h(f)$, the controls noise dependent \ac{SNR} becomes
\begin{align}\label{eq:SNR based on controls noise}
  \mathrm{SNR}^2 = \mathrm{SNR}_{\text{det}}^2 - \mathrm{SNR}^2_{\text{lost}}\text{,}
\end{align}
where
\begin{align}\label{eq:SNR^2_det}
  \mathrm{SNR}_{\text{det}}^2 = 4\int_{0}^{\infty} \frac{\left | h_{\mathrm{gw}}(f)\right |^2}{S_{\mathrm{det}}(f)} \ df
\end{align}
and
\begin{align}\label{eq:SNR^2_lost}
  \mathrm{SNR}_{\text{lost}}^2 = 4\int_{0}^{\infty} \frac{\left |C\right |^2\left | h_{\mathrm{gw}}(f)\right |^2 }{S_{\mathrm{det}}(f)^2}  S_{\mathrm{c}}(f) \ df \mathrm{.}
\end{align}
Adapting Eq.~\ref{eq:d_range} gives the controls-noise-dependent expression for the detection range lost as
\begin{align}\label{eq:detection range lost}
  d_{\mathrm{lost}} =\frac{d_{\mathrm{gw}}}{2.26\rho_0} \frac{\mathrm{SNR}^2_{\text{lost}}}{2\mathrm{SNR}_{\text{det}}}\mathrm{.}
\end{align}
The lost detection range can also be computed using the linearization. The point of linearizing the equation is to split it into terms that show the specific impact of a particular noise, $S_{\mathrm{c}}$, on the \ac{SNR} and \ac{BNS} range. In this equation, $S_{\mathrm{c}}$ can be any noise of interest, assuming it is small enough to keep the linearization accurate. The $S_{\mathrm{c}}$ term is inside an integral that includes a frequency-dependent weighting. That weighting term will later be useful to define a shaping filter for the controls optimization problem in terms of the lost \ac{SNR} or range cost.

The \ac{RMS} of a noise from channel $N_{\mathrm{c}}$, is indicated by $\rms_{\mathrm{c}}$. In order to keep the interferometer locked, there is a maximum allowable \ac{RMS}. If the noise in the interferometer exceeds that limit, the interferometer is unable to maintain operation, e.g., sufficient alignment.
\begin{align}
  \rms_{\mathrm{c}} \equiv \int_{0}^{\infty} S_{\mathrm{c}} (f) df <& \rms_{\mathrm{c, unlock}} \mathrm{.}
\label{eq:RMS_unlock}
\end{align}
A fundamental requirement of control systems is to ensure that no \ac{BNS} exceeds its lock-loss threshold. Thus, there is a tradeoff between operational requirements and noise injection.

There are two primary sources of controls noise: environmental noise and measurement noise. Environmental noise is any noise that acts on the states of the suspended interferometer plant. For example, seismic noise is an environmental noise that is of great concern for ground-based interferometers. Measurement noise is the difference between the measured states of a system and their true value, generally due to an imperfect intermediate sensor used to probe the states. Both the measurement and environmental noise pass through a control filter and then feed back into the system, but measurement noise would not otherwise affect plant states except through the control system actuating it into real plant noise.

Considering only two noise sources is a useful simplification. In reality, there are many noise sources that all contribute differently to every \ac{DOF} in the interferometer. However, these noise sources can only enter the system in two places: either before the plant (environmental noise) or before the controller (measurement noise).
This distinction also aligns with the assumptions and model of a typical optimization using classical control theory to shape the open-loop gain.

Without physically decreasing environmental or measurement noise, which much of the engineering resources \ac{LIGO} is dedicated to, the only way to decrease controls noise is to find an optimal controller, \controller, such that the minimum amount of noise is fed through into \ac{DARM} while achieving a total \ac{RMS} goal.

Here we distinguish two channels $N_{\text{p}}$ and $N_{\text{a}}$. The former, which must meet a noise threshold, is the sum of environmental noise and controller actuation, filtered by the plant's transfer function. This noise, $N_{\text{p}}$, is the noise directly seen by the interferometer. The latter, $N_{\text{a}}$, is the signal sent by an actuator due to its control system, then shaped by the plant's transfer function, without including direct contributions from environmental noise. These two noises are soon presented visually in Fig.~\ref{fig:systemlayout}. The optimal controller must ensure that the plant's noise $\rms_{\mathrm{p}}$ will meet requirements by shaping its spectrum $S_{\mathrm{p}}$ through the environmental and measurement noise contributions.
\begin{align}
S_{\mathrm{p}}(f) &= \left |\frac{P(f)}{1-G(f)}\right |^2  S_{\mathrm{env}}(f) + \left |\frac{G(f)}{1-G(f)}\right |^2  S_{\mathrm{meas}}(f) \text{,} %closed loop
%   &&&
% N_{\mathrm{p}}(f) &= \frac{P(f)}{1-G(f)}  N_{\mathrm{env}}(f) + \frac{G(f)}{1-G(f)}  N_{\mathrm{meas}}(f) \text{.} %closed loop
\label{eq:CLControlsnoise}
\end{align}
where
\begin{align}
  G(f)&=K(f)P(f)
\label{eq: G=KP}
\end{align}
is the open-loop gain of the full system, $P(f)$ is the transfer function of the plant component, either the input or output of $P(f)$ represents the degree of freedom to be controlled, $K(f)$ is the transfer function of the controller, $S_{\mathrm{env}}(f)$ is the \ac{PSD} of the environmental noise, and $S_{\mathrm{meas}}(f)$ is the \ac{PSD} of the measurement noise. Here, it is explicitly indicated that each of these expressions is a function of frequency; this dependence is continued but will not be explicitly stated in the following formulas. 

The total plant noise, $N_{\text{p}}$, which is presented in Eq.~\ref{eq:CLControlsnoise}, is the typical minimization metric for modern controls. It is the measure of all the noise coming out of a system, including all coherent cancellations and interactions. In a system with no cross-coupling to any other degree of freedom, minimizing $S_{\mathrm{p}}$ would minimize the sum of both measurement noise and environment noise injected into the system.

% By setting the $K=0$, a condition that should eliminate all controls noise, $S_{\mathrm{p}} = \left |P\right |^2 S_{\mathrm{meas}}$ meaning that in the $K\neq0$ case, $S_{\mathrm{p}}$ is actually a combination of measurement noise and the environmental noise filtered by the plant. While $S_{\mathrm{p}}$ is being called the \ac{PSD} of the controls noise, it also includes the pure environmental noise. It may seem counterintuitive to include environmental noise in a definition of controls noise, but since the goal of control systems is to minimize the impact of all noise, it provides a single useful metric.

The definition of $N_{\text{p}}$ is often presented as controls noise in modern controls. However, this differs from most previous work in gravitational wave interferometer sensing and control, where controls noise is defined as any noise that passes through the controller and becomes a new contribution through actuation on the \ac{DOF}. The definition of actuation point controls noise, $S_{\mathrm{a}}$, is 
\begin{align}
  S_{\mathrm{a}} &= \left |\frac{PG}{1-G}\right |^2  S_{\mathrm{env}} + \left |\frac{G}{1-G}\right |^2  S_{\mathrm{meas}}\text{,}
  \label{eq:CLActuationControlsnoise S_a}
\end{align}
and
\begin{align}
  N_{\mathrm{a}} &= \frac{PG}{1-G} N_{\mathrm{env}} +\frac{G}{1-G} N_{\mathrm{meas}}
  \text{.} %closed loop
\label{eq:CLActuationControlsnoise N_a}
\end{align}
The actuation point controls noise definition is useful because optimizing to reduce it is more directly a means of penalizing actuation, as $K$ is in the numerator of both environmental and measurement noise contributions to Eqs.~\ref{eq:CLActuationControlsnoise S_a} and~\ref{eq:CLActuationControlsnoise N_a}. In large multi-\ac{DOF} coupled systems like \ac{LIGO}, it is often true that actuation on one \ac{DOF} can inject noise into another. The \ac{PSD} $S_{\mathrm{c}}$, from Eq.~\ref{eq:SNR based on controls noise}, could be set equal to any noise of interest like $S_{\mathrm{p}}$; however in this paper $S_{\mathrm{c}}=S_{\mathrm{a}}$ as is appropriate for the alignment-control \ac{DOF} we investigate and as typical for \ac{GW} controls. The noise from this metric, $N_{\mathrm{a}}$, is then used as the noise, $N_{\mathrm{a}}$, in Eq.~\ref{eq: h_gw}. An approach more widely used in modern controls that only uses $S_{\mathrm{p}}$ as an optimization metric is discussed in Appendix~\ref{sec:alternate_FOM}.

In this work, both $S_{\mathrm{a}}$ and $S_{\mathrm{p}}$ will be used. The former for the control effect on SNR and BNS range, and the latter for the minimization of the total \ac{RMS}, $\rms_{\mathrm{p}}$ from Eq.~\ref{eq:RMS_unlock}, needed to maintain the operating point. Finding a ``best'' controller is difficult because of the trade-off between suppressing environmental noise in $S_{\mathrm{p}}$ with a large $K$ and suppressing measurement noise in $S_{\mathrm{a}}$ with a small $K$. Finding the optimal controller requires a multi-objective optimization: the \ac{RMS} controls noise, $\rms_{\mathrm{p}}$, and the \ac{BNS} detection range.

\subsection{Application to Alignment Sensing Control System}
\label{subsec:}
Controls noise associated with the \ac{ASC} system is one of the largest sources of controls noise at low frequencies~\cite{O3sensitivity, O4SensitivityCapote}. The \ac{ASC} system is responsible for keeping the \ac{LIGO} test masses aligned in their angular degrees of freedom. Each \ac{LIGO} interferometer has eight principal control filters dedicated to maintaining angular stability in the arm cavities. These eight control filters are the product of three distinguishing degrees. Each system is either in pitch or yaw (P/Y) rigid-body rotation, Differential or Common motion between the two Michelson arms (C/D), and either HARD or SOFT spring reaction to motion.

HARD modes are when the two test masses that make up \ac{LIGO}'s Fabry-P\'erot arm cavities rotate in opposite directions, e.g., one rotates very slightly up, and the other rotates very slightly down. Since \ac{LIGO}'s Fabry-P\'erot cavities have a high g-factor, the differential rotation causes a rotation of the optical axis, and radiation pressure pushes the arms back towards their nominal angle. Thus, HARD modes increase suspension stiffness and occur at higher frequencies. SOFT modes are the opposite, where the mirrors move at a common angle, which translates the cavity axis. The translation moves the beam off center and produces a negative spring constant (anti-spring).

The name of each mode is the composition of these distinguishing degrees. I.e., CSOFTY is the common soft mode in the yaw angular tilt. In this study, only the DHARD yaw controller is calculated. The choice of actuation point noise optimization for the DHARD yaw \ac{DOF} presented in this paper was motivated by the strong coupling between angular actuation and length in \ac{LIGO} Hanford's arms.

\section{Problem Formulation}
\label{sec:Problem Formulation} 
The heart of the controls noise problem is straightforward to outline: too much measurement noise gets passed through the controller and is actuated spuriously into the \ac{DARM} \ac{DOF}, injecting noise. As with all \ac{LIGO} systems, the goal is to minimize the noise that is fed into the readouts of the interferometer that contain information of astrophysical importance, mainly the \ac{DARM} readout. The solution to this control problem is more difficult, as it must account for many different factors, including the plant's response, the measurement noise passed through the control system, and the system's stability and robustness.

The modern controls optimization method that is well-suited to this problem is \ac{LQG} control. An \ac{LQG} controller is the optimal linear controller for a system that is disturbed by Gaussian white noise and has a quadratic (\ac{RMS}-like) cost function. Described in Sections~\ref{subsec:calcH2norm} and~\ref{subsec:optController}, \ac{LQG} controllers minimize a system's $\mathcal{H}_2$ norm, which is a measure of the \ac{RMS} at system outputs from all of its noise inputs. The ``$\mathcal{H}$'' stands for solutions within a Hardy space, indicating that the system must be closed-loop stable. This is necessary to have well-posed noise in the time domain, and appropriate for implementable feedback controllers. The typical quadratic cost function for LQG is based on the total noise \ac{RMS}, $S_\text{p}$, of a \ac{DOF}. \ac{LIGO}'s performance relies on multiple criteria, so we will use a weighted \ac{RMS} that includes both the total noise and the lost detection range as figures of merit.

So far, the noise model of Section~\ref{subsec: controls noise} has not included the nonlinear/bilinear aspects. It has simply been a model of direct, linear noise coupling, which should naively only be optimized for the \ac{BNS} range while ignoring the \ac{DOF} \ac{RMS}. The following sections investigate how to incorporate a weighted \ac{RMS} to balance noise suppression against control noise injection.

\subsection{System Overview}
\label{subsec:SystemOverview}

\begin{figure*}[!t]
  \centering
  \includegraphics[width=0.8\linewidth]{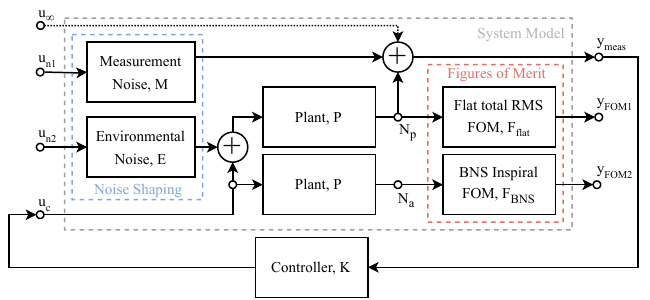}
  % \autographicsdrawio{
  %   folder=./Graphics/,
  %   width=0.8\linewidth,
  %   file=ASCSimpSystemOverview_F3_withP,
  % }
  \caption{The layout of the modeled system with a controller attached. Here, the input white noise, $u_{n1}$, is being shaped to appear as environmental noise seen by the plant. The system has two \acp{FOM}, which are simply transfer functions that weigh the output noise: a broadband \ac{FOM} outputs the unshaped noise, i.e., every frequency has the same weight, and a \ac{BNS} inspiral \ac{FOM}, which represents the band of detection for \ac{BNS} signals. In this model, the two plant blocks are identical and serve as a way to represent $S_{\mathrm{a}}$ for actuation-point controls noise minimization. These \acp{FOM} are discussed further in Section~\ref{subsec:foms}. The input associated with $u_{\infty}$, which has a dotted line, is used for the $\mathcal{H}_\infty$ minimization discussed in Section~\ref{subsec: h_inf simple explination}.}
  \label{fig:systemlayout}
\end{figure*}

When creating a \ac{SISO} controller for a single \ac{DOF}, a \ac{MIMO} system is constructed that takes into account the multiple inputs that can affect the \ac{DOF}, as well as multiple outputs to measure its performance. This system, whose block diagram is presented in Fig.~\ref{fig:systemlayout}, includes all of the information that is used to find the optimal controller, mainly of three types: the noise in the system, the plant of the system, and the \acp{FOM} of the system. The first two of these pieces are rather straightforward and regularly measured at the \ac{LIGO} sites, while the motivation for the \acp{FOM} is given in Section~\ref{subsec:foms}. 

The full system model, outlined by the dashed grey box in Fig.~\ref{fig:systemlayout}, includes four inputs, a control input, $u_{c}$, a measurement noise input, $u_{n1}$, a state noise input (environmental noise), $u_{n2}$, and an $\mathcal{H}_\infty$ disturbance input. Both noise inputs are white noise inputs that pass through shaping filters to model the measurement and environmental noise. The control input,$u_{\mathrm{c}}$ is connected to the output of the controller and is not a white noise input. The $\mathcal{H}_\infty$ disturbance input, $u_\infty$, serves as a port for the solver to inject unshaped signals into the system and is a necessity for optimal control analyzed later. The system has two types of outputs: the figure of merit outputs and the measured output, $y_{\text{meas}}$. $y_{\text{meas}}$ gives the state of the plant with added measurement noise. Measurement noise is only observed by the controller and represents an imperfect estimate of the plant's true state.

\begin{figure}[!ht]
     \centering
     \includegraphics[width=1\linewidth]{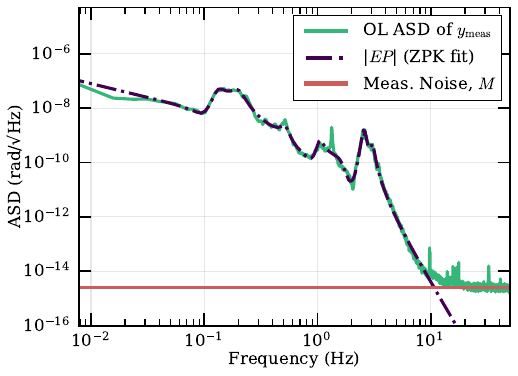}
     \caption{
       The noise spectrum of the DHARD yaw \ac{DOF} at the \ac{LIGO} Hanford Observatory decomposed into fits of measurement and environmental contributions.
       The spectrum are fit to filters $M$ and $E$, see relations of Eqs.~\ref{eq:EM_relations S_Env}-\ref{eq:EM_relations S_un1}.
       The measured ASD of $y_\text{meas}$ is the measured spectrum of the cavity axis (see Fig.~\ref{fig:systemlayout}) and is equal to $|EP+M|$. This spectrum assumes no control, $K=0$ (open-loop). After fitting a ZPK to $|EP|$, the contribution from $E$ was isolated by dividing by the model for the plant ($P$).
       The filters are using the Python package \texttt{wield.iirrational} to refine fits with gradient descent after making initial guesses using the AAA algorithm~\cite{iirrational, NakatsukasaSJSC18AAAAlgorithm}.}
     \label{fig:ASCfit}
\end{figure}

The system layout includes multiple blocks, each representing a transfer function. The plant, $P$, represents the physical system being simulated; in this case, it is the mechanism by which the actuator converts force/torque inputs into displacements/tilts. The measurement noise, $M$, is the shaping filter that turns the input Gaussian white noise into the appropriately shaped sensing noise, while the environmental noise, $E$, serves the same purpose for the state noise on the plant, both shown in Fig.~\ref{fig:ASCfit}.
The transfer functions relate to Eqs.~\ref{eq:CLControlsnoise},~\ref{eq:CLActuationControlsnoise S_a}, and~\ref{eq:CLActuationControlsnoise N_a} through
\begin{align} \label{eq:EM_relations S_Env}
S_{\mathrm{env}} &= \left |E\right |^2 S^u_{n1}
\end{align}
and
\begin{align} \label{eq:EM_relations S_meas}
S_{\mathrm{meas}} &= \left |M\right |^2 S^u_{n2} \mathrm{,}
\end{align}
where
\begin{align} \label{eq:EM_relations S_un1}
   S^u_{n1}\text{, }  S^u_{n2} &\equiv 1/\sqrt{\mathrm{Hz}}
\end{align}
and are uncorrelated. This indicates that these inputs are driven by Gaussian white noise that is then shaped by the relevant noise-shaping filters. The two \ac{FOM} filters, $F_{\mathrm{BNS}}$ and $F_{\mathrm{flat}}$, represent the transfer function that weights the frequency domain signal properly for two different measures of the controller's performance described in Section~\ref{subsec:foms}. 

%\LM{quick callout that each FOM is needed to vary between $S_p$ and $S_a$ weights for $S_c$}
%\LM{Needs to point out the connections to $N_a$ $N_b$}

The two \ac{FOM} outputs are free of any direct measurement noise, and they are each connected to a point representing the $N_{\text{p}}$, and $N_{\text{a}}$ plant and actuator noise terms of Eqs.~\ref{eq:CLActuationControlsnoise S_a},~\ref{eq:CLActuationControlsnoise N_a}, and~\ref{eq:CLControlsnoise}. The frequency response of the plant must have two copies in our model to produce the different contributions of $K$ (to be designed) in the expressions of $N_{\mathrm{a}}$ and $N_{\mathrm{p}}$.

The Figure of Merit block includes a shaping filters for each path that weight the relative contribution of $S_a$ or $S_p$ in the optimization problem. The construction and weighting of the \ac{FOM} filters is described in Section~\ref{subsec:foms} and is fundamental to finding the Pareto front of optimal controllers for a given system.
The absence of measurement noise indicates that the \ac{FOM} outputs report the true value (up to weighting) of the \ac{DOF} from which it is constructed and not what would be observed through an imperfect physical measurement.
%An example of this is trying to use controls to move a car to a specific position using \ac{GPS}. The position of the car is a real set of two coordinates, but the controller moving the car only has the information from the noisy \ac{GPS} position. If one were to quantify how well the controller was moving the car around, one would want to compare the real position of the car to the desired position. In this example, the measured output is the \ac{GPS} signal, and the \ac{FOM} output is the difference between the real position of the car and the desired position.

For this alignment control application, the state of the plant at $N_{\mathrm{p}}$ represents the total angular motion of the mirror or cavity in its degree of freedom (Differential, HARD, Yaw). A flat filter representing the first \ac{FOM} then has $y_{_{\mathrm{FOM1}}}$, which integrates into the total variance of the mirror motion. This is the typical objective that is minimized in LQG/$\mathcal{H}2$ optimal control. The $N_{\mathrm{a}}$ connection to $y_{_{\mathrm{FOM2}}}$ represents how the control actuation pushes on the mirror in pitch or yaw. If the beam is miscentered, the alignment actuation is converted into length noise. The BNS FOM then contains a weighing filter for how important this length noise conversion is for the SNR or range figure of merit.

\subsection{Figures of Merit}
\label{subsec:foms}
As discussed in the previous sections, \acp{FOM} are measures of the performance of a closed-loop system. In the example presented in this paper, an essential new complexity is added: multiple \acp{FOM}.

\begin{figure}
     \centering
     \includegraphics[width=1\linewidth]{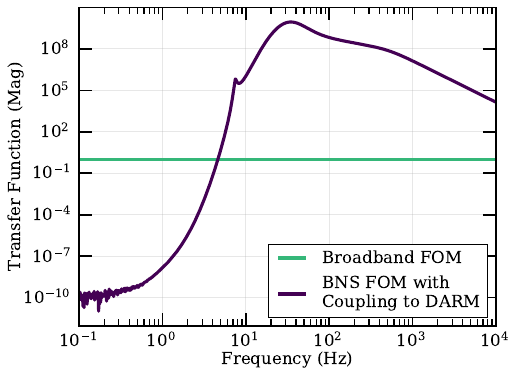}
     \caption{
       The magnitude of two \ac{FOM} filters used in the augmented system, as shown in Fig.~\ref{fig:systemlayout}. Both of these \acp{FOM} shape either $N_{\mathrm{p}}$ or $N_{\mathrm{a}}$ to create metrics for the controller to minimize. The broadband \ac{FOM} results in an unshaped signal that must be below a specified threshold to keep the interferometer locked. The \ac{BNS} \ac{FOM} shapes $S_{\mathrm{a}}$ allowing for the direct optimization of \ac{LIGO}'s BNS detection range.}
     \label{fig:FOMfit}
\end{figure}

In the model of the system, shown in Fig.~\ref{fig:systemlayout}, two \acp{FOM} are shown, which have been indicated by Eqs.~\ref{eq:SNR based on controls noise} and~\ref{eq:RMS_unlock}. Shown inside the red dashed outline in Fig.~\ref{fig:systemlayout} and represented by transfer functions, these two \acp{FOM}, $F_{\mathrm{BNS}}$ and $F_{\mathrm{flat}}$, allow for different frequency weightings of the output signal. Fig.~\ref{fig:FOMfit} shows the magnitude of the two \acp{FOM} used in the augmented \ac{ASC} system.
The first broadband \ac{FOM}, $F_{\mathrm{flat}}$, is required to minimize $\rms_{\mathrm{p}}$ which is necessary to preserve the inequality in Eq.~\ref{eq:RMS_unlock}. The broadband \ac{FOM} represents the unshaped total noise in the \ac{DOF} seen after the plant, which affects the operational locking of the interferometer. This \ac{FOM} is also necessary in order for the problem to be well-posed for optimal control, ensuring Assumption~\ref{itm:nonsingularD12} in Section~\ref{subsec:optController}.

The second \ac{FOM}, passing through $F_{\text{BNS}}$, is a measure of how much the alignment actuation influences the \ac{SNR} or range of the \ac{LIGO} detector to an inspiral of two $1.4~M_\odot$ neutron stars at a distance of 40~Mpc, using Eqs.~\ref{eq:SNR based on controls noise}-\ref{eq:SNR^2_lost}. The inspiral determines the function $h_{\text{gw}(f)}$. The factor $C(f)$ is a coupling factor for actuation point control noise into DARM, computed using noise budget data from~\cite{O4SensitivityCapote}. $S_{\mathrm{det}}$ is the detector's noise \ac{PSD}. We fit the weighting filter so that
\begin{align}
\left| F_{\text{BNS}}(f) \right| = 2\frac{\left |C(f)\right |\left | h_{\mathrm{gw}}(f)\right | }{S_{\mathrm{det}}(f)}
  \cdot
\frac{d_{\mathrm{gw}}}{16\cdot 2.26 \cdot \mathrm{SNR}_{\text{det}}}\text{,}
\end{align}
where the first term creates a weight filter to compute the $\text{SNR}_{\text{lost}}$, and the second term converts to lost range.

By connecting to the $N_{\mathrm{a}}$ actuation-point noise, the total variance of $y_{_{\mathrm{FOM2}}}$ then produces the lost SNR or lost range. Practically, this filter (seen in Fig.~\ref{fig:FOMfit}) has a dynamic range going from low to high frequencies that spans more than the 16 orders of magnitude of a double-precision float, creating issues that we detail in the following sections.

The frequency-dependent coupling function $C(f)$ is measured through an excitation injected into the control loop~\cite{O4SensitivityCapote}. When excited, $S_\mathrm{a}$ from Eq.~\ref{eq:CLActuationControlsnoise S_a} becomes
\begin{align}
S_{\mathrm{a}\text{, exc}} &= S_\mathrm{a} + \left |\frac{G}{1-G}\right |^2  S_\mathrm{exc}\mathrm{,}
\label{eq: S_P excited}
\end{align}
where $S_\mathrm{exc}$ is the \ac{PSD} of the excited signal summed in before the controller. When this excitement is added, $S_h$ from Eq.~\ref{eq:S_h} becomes
\begin{align}
 S_{h\text{, exc}} &= S_{\mathrm{det}} + |C|^2 \left (S_\mathrm{a} + \left |\frac{G}{1-G}\right |^2  S_\mathrm{exc} \right ) \mathrm{.}
 \label{eq: S_h excited}
\end{align}
From Eqs.~\ref{eq: S_P excited} and~\ref{eq: S_h excited}, $|C|^2$ can be calculated and fit from
\begin{align}
|C|^2 = \frac{S_{h\text{, exc}} - S_h}{S_{\mathrm{a}\text{, exc}} - S_\mathrm{a}}\mathrm{.}
\label{eq:A2L coupling}
\end{align}

\subsubsection{Pareto Front of Multiple FOMs for Bilinear Noise}

To fully motivate the use of these two \acp{FOM} in the bilinear noise case, Eq.~\ref{eq:SNR based on controls noise} can be rewritten to directly depend on both \acp{FOM}. Consider the noise coupling factor $C$ from Eq.~\ref{eq:SNR^2_lost}. For the multiplication effect that occurs for bilinear noise on a single \ac{DOF}, a model is assumed where $C \propto \rms_{\mathrm{p}}$ represents an intermodulation effect. The effect of an intermodulation can be seen by rewriting Eq.~\ref{eq: h_gw} in the time domain, with the control noise term becoming bilinear
\begin{align}
 h_\mathrm{signal}(t) &= h_{\mathrm{gw}}(t) + N_{\mathrm{det}}(t) + \chi N_{\mathrm{a}}(t)N_{\mathrm{p}}(t)\mathrm{.}
 \label{eq: h_gw bilinear}
\end{align}
Where $\chi$ is a scale factor for the cross-coupling. For our alignment example, $N_{\mathrm{p}}$ is the residual angular motion of the mirror, while $N_{\mathrm{a}}$ is the actuated tilt. If the residual angular motion is non-zero, then the applied tilt causes a coupling to arm length.

When the $N_{\mathrm{p}}$ noise is dominated by low-frequency components, this term becomes $\chi \rms_{\mathrm{p}} N_{\mathrm{a}}(t)$ by linearizing. See Appendix~\ref{sec:bilinear_noise_spectrum} for a formal derivation. This substitution can then be used in Eq.~\ref{eq:SNR^2_lost} via  $C(f) = \chi(f) \cdot \rms_{\mathrm{p}}$. In the alignment control example problem, this relation for constructing $C(f)$ yields the coupling factor called ``A2L'' for alignment-to-length. This factor is measured~\cite{O4SensitivityCapote} and translates angular motion in the alignment system into length motion in the \ac{DARM} gravitational wave channel. This form for $C$ and similar models of \ac{RMS}-dependent coupling factors indicates that one aspect of the range depends on the \ac{RMS} as computed from the flat \ac{FOM}, while the other depends on the \ac{RMS} as computed from the shaped \ac{BNS} \ac{FOM}.

Assuming this form of the $C$ coupling function, the \ac{SNR} loss of Eq.~\ref{eq:SNR^2_lost} becomes
\begin{align}
  \mathrm{SNR}_{\mathrm{lost}}^2 &= 4 \cdot \rms^2_{\mathrm{p}} \cdot \int_{0}^{\infty} \frac{\left |\chi\right |^2\left |h_\mathrm{gw}(f)\right |^2 }{S^2_{\mathrm{det}}(f)}  S_{\mathrm{c}}(f) \ \rd\!f 
   \\
   &=
     \underbrace{\int_{0}^{\infty} \left| F_{\mathrm{flat}} \right|^2 S_{\mathrm{p}} \ df}_{\rms^2_{\mathrm{flat}}} \cdot \underbrace{\int_{0}^{\infty} \left| F_{\mathrm{BNS}} \right|^2 S_{\mathrm{c}} \ \rd\! f}_{\rms^2_{\mathrm{BNS}} / \overline{R}^2_{\mathrm{flat}}}
   \mathrm{,}
   \label{eq:SNR lost}
\end{align}
where $\rms^2_{\mathrm{flat}}$ is the RMS of the total plant motion during control and $\overline{R}^2_{\mathrm{flat}}$ is a constant for RMS at the time of the measurement of the A2L coupling factor $C(f)$.

As a minimization problem, a necessary condition for optimality can be established by taking the derivative with respect to the controller parameters $\controllerSS$, in bold to represent parameters in a state-space matrix form. Setting these derivatives to zero finds an extremal point of the noise as
\begin{align}
  0 &= \frac{1}{4|\chi|^2} \frac{\rd}{\rd \controllerSS} \mathrm{SNR}_{\mathrm{lost}}^2\\
  &=
    \rms^2_{\mathrm{flat}}
    \frac{\rd}{\rd \controllerSS} \rms^2_{\mathrm{BNS}} + 
    \rms^2_{\mathrm{BNS}}
    \frac{\rd}{\rd \controllerSS} \rms^2_{\mathrm{flat}}
   \mathrm{.}
   \label{eq:SNR lost use}
\end{align}
This relation can be satisfied through the equivalence
\begin{align}
  0=\frac{\rd}{\rd \controllerSS} \rms^2_{\mathrm{syn}}\text{,}
  \label{eq:SNR lost equiv}
\end{align}
when
\begin{align}
    \zeta &\equiv \frac{\rms_{\mathrm{BNS}} }{\rms_{\mathrm{flat}}}\mathrm{.}
\end{align}
The \ac{LQG} and mixed-sensitivity approaches below find control systems that meet this necessary condition. By scanning over the weight parameter $\zeta$, the Pareto front of optimal controllers is produced. The above derivation indicates that the bilinear-optimal controller exists along this Pareto front.

There are multiple reasons why the parameter $\zeta$ should be abstracted as an independent term for Eq.~\ref{eq: RMS_syn}. The main reason is that, prior to optimization, the ratio of the \ac{RMS} values of the two \acp{FOM} cannot be known. Thus, there is no presumptive $\zeta$ to target. Since \ac{LQG} can only optimize over a single parameter, all of a system's \acp{FOM} must be combined. With $\zeta$, we can define a synthetic weighted \ac{RMS} of the two \acp{FOM}, which meaningfully relates to the above condition on the derivative of the \ac{SNR}.
\begin{align}
    \rms_{\mathrm{syn}}^2 = \zeta^2 \rms_{F_{\mathrm{BNS}}}^2 + \rms_{F_{\mathrm{flat}}}^2 \mathrm{,}
    \label{eq: RMS_syn}
\end{align}
The weighted sum of the two \ac{FOM} outputs creates a single synthetic output with its own \ac{RMS}, $\rms_{\mathrm{syn}}$. The problem now boils down to a problem to minimize $\rms_{\mathrm{syn}}$ over many given weightings of $\zeta$. It is advantageous to only weight one of the \acp{FOM} to achieve the desired relative weighting because leaving the weighting of the flat \ac{FOM} untouched allows for consistent $\mathcal{H}_\infty$ norms across different weightings, which will be important in Section~\ref{subsec: h_inf simple explination}.

For values of $\zeta$ that are zero, or very large, the optimization of $\rms_{\mathrm{syn}}$ will find the minimum \ac{RMS} in one of the respective \acp{FOM}. Because the \ac{RMS} values satisfy a minimization constraint, the optimized values of both $\rms_{\mathrm{BNS}}$ and  $\rms_{\mathrm{flat}}$ are expected to be monotonic but not necessarily smooth or convex in $\zeta$. 

Together, this suggests that scanning $\zeta$ will allow one to find the global minimum of the lost \ac{SNR}, but doesn't establish that the control-optimal \ac{SNR} loss function is a convex optimization problem in $\zeta$. For any more complex dependence of $C$ on $\chi$ or $\rms_{c}$, the derivative argument above is expected to still apply, but the relationship of Eq.~\ref{eq:SNR lost equiv} will become more complex. Finally, the above derivation and example based on the $C$ coupling parameter assume only a single \ac{DOF}. More elaborate models of the $C$ parameter are likely dependent on the \ac{RMS} of multiple \acp{DOF}. Thus, the decision point of where to balance flat \ac{RMS} vs. the \ac{BNS}-weighted \ac{RMS} is, in general, more complex. These are major reasons that $\zeta$ should be treated abstractly to produce a Pareto front. The following plots show that the $\zeta$ parametrization clearly shows the boundary of the optimal space for the two major \ac{RMS} contributions. Thus, scanning $\zeta$ to find a boundary that shows the specific tradeoff between flat-weighted \ac{RMS} and the \ac{BNS}-weighed \ac{RMS} is the most generally useful expression and is achieved through the $\zeta$ parameter. We believe the above argument can apply to \ac{MIMO} systems using a vector of $\zeta$ scales, but leave this for future work. 

\section{LQG Optimization}
\label{sec:LQG Solution Simple}

The \ac{LQG} problem is to find an optimal controller that yields the lowest $\rms_{\mathrm{syn}}$ for a given $\zeta$. This is equivalent to finding a controller that minimizes the $\mathcal{H}_2$ norm of a \emph{transfer-function-matrix}, $\mathbf{T}_2(f)$, from the $u_{n1}, u_{n2}$ vector of inputs, to the $y_{_{\mathrm{FOM1}}}, y_{_{\mathrm{FOM2}}}$ vector of outputs, with a $\zeta$ scaling first applied to the $y_{_{\mathrm{FOM2}}}$ output.

By contrast, \ac{LQG} is often taught and presented starting from the separation principle, combining a Kalman filter as a noise-optimal state linear-quadratic estimator with a linear quadratic regulator that minimizes an actuator cost function. The actuator cost is often described as fuel consumption, but when using the \ac{FOM} weighting filters, it can also be interpreted as part of a noise-minimization procedure. This interpretation of \ac{LQG} as an $\mathcal{H}_2$ noise minimization procedure can be established intrinsically through minimization~\cite{AthansITAC67DirectDerivation}, as well as through the case analysis of the separation principle~\cite{DoyleITAC89StatespaceSolutions}. 

The $\mathcal{H}_2$ optimization can prove difficult by hand or even with a gradient-descent numerical optimizer because of design trade-offs. Finding the perfect balance between suppressing environmental noise and introducing minimal measurement noise, all while weighting the output by two \acp{FOM}, is a job that optimal control can perform in a single algebraic solution step. Here,\ac{LQG} is used to find the optimal controller to give the lowest possible $\rms_{\mathrm{syn}}$. The control transfer function, $K$, is determined by the state-space description of the control, $\mathbf{K}_{\mathrm{LQG}}$, using the method described in Section~\ref{subsec:optController}. The method of the numerical solver is not essential to understanding the results of this paper, but an important point is that \ac{LQG} optimization is non-iterative. This is essential for speed, as our method requires a scan over the $\zeta$ parameter that indicates the tradeoff between two \acp{FOM}. Because of dynamic range and numerical challenges, an understanding of the numerical solver has been essential to implement this work with a reliable software implementation, but ideally, it should not be required for end users.

\subsection{LQG Optimal Controllers}
\label{subsec:LQG Results Simple}
Using the plant system, whose layout is presented in Fig.~\ref{fig:systemlayout}, a set of optimal \ac{LQG} controller filters was calculated using the method described in Section~\ref{subsec:optController}. These controllers, corresponding to filter block $K$ in Fig.~\ref{fig:systemlayout}, were calculated for different weights of the two \acp{FOM} corresponding to the \ac{BNS} range and total \ac{RMS} for a range of $\zeta$ values. The change in the weighting $\zeta$ causes the \ac{BNS} \ac{FOM} to increase its contribution to the $\rms_{\mathrm{syn}}$.
By calculating a number of the $\zeta$ values, it is possible to show the performance of those controllers in relation to the \ac{LIGO}'s early observing run O4 DHARDY open-loop gain with respect to any tradeoff between total noise and lost performance (observing range). Since the solver is always finding the lowest noise across different weightings, the results form a Pareto front, showing the boundary of all optimal linear controllers for this cavity-alignment degree of freedom.

\begin{figure*}
\centering
\includegraphics[width=0.78\linewidth]{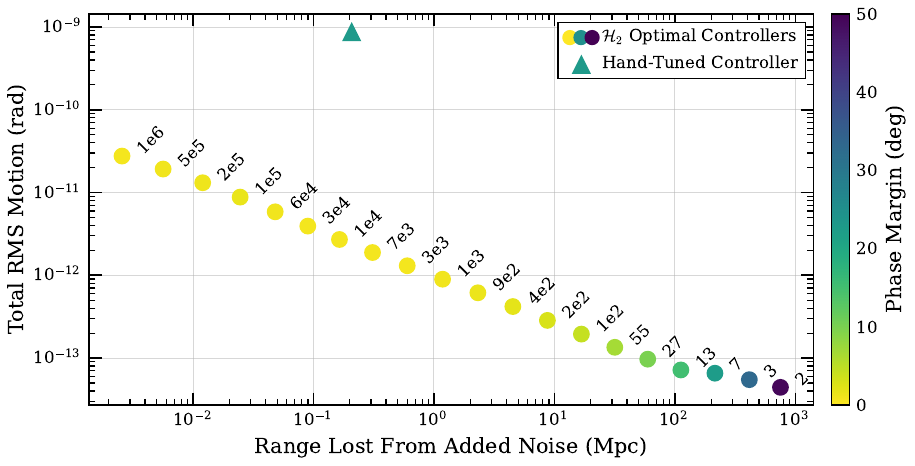}
\caption{The \ac{RMS} noise at the two \ac{FOM} outputs given the white noise disturbances for the closed-loop system, including the system shown in Fig.~\ref{fig:systemlayout} and optimal \ac{LQG} controllers calculated for specified $\zeta$ values. Each point represents the performance of a single controller that is the \ac{LQG} optimal solution for the plant with a different \ac{FOM} weighting. The \ac{RMS} value is shown on the axis of the corresponding \ac{FOM}. The \ac{BNS} \ac{FOM}'s \ac{RMS} is calibrated to represent \ac{BNS} range lost by the noise from a given controller, while the flat \ac{FOM} represents the \ac{RMS} noise in radians for \ac{ASC} DHARD yaw. The number next to each point represents the $\zeta$ value used to acquire the desired weighting. The color of each point corresponds to the phase margin of the controller. The triangle represents the current \ac{LIGO} hand-tuned controller's performance and phase margin. All of the \ac{LQG} optimal controllers whose performance beats the hand-tuned controller in both \acp{FOM} have poor phase margins, and they get worse as the \ac{BNS} range \ac{FOM} is emphasized using a higher $\zeta$ factor. The margins can be improved by the \ac{LQG} optimal solutions with an $\mathcal{H}_\infty$ performance bound discussed in Section~\ref{sec: LQG with arb margins} and presented in Fig.~\ref{fig:RMSplot}.}

\label{fig:RMSplot H2}
\end{figure*}

Fig.~\ref{fig:RMSplot H2}, shows a comparison between several \ac{LQG} optimal controllers for several $\zeta$ values. To calculate the controllers, a system scaled by $\zeta$, is used. During the solve, if $\zeta$ is too large, the solver will either fail or create sub-optimal controllers due to numerical error in eigenvalue computation and loss of the ability to order deflating subspaces. Once the controller is calculated, it is connected to an unscaled system, where $\zeta=1$. The unscaled system models the controllers' performance, as a Lyapunov equation can compute the lost BNS range by computing the RMS of the BNS FOM output. The two performance metrics are then plotted. Lower is better, representing either suppressed noise in the alignment plant or less loss of detection range. The \ac{LQG} open-loop gain of the controllers is shown in comparison to a hand-tuned \ac{LIGO} loop design for DHARD yaw used during LIGO's fourth observing run.

\begin{figure*}
\centering
\includegraphics[width=0.7\linewidth]{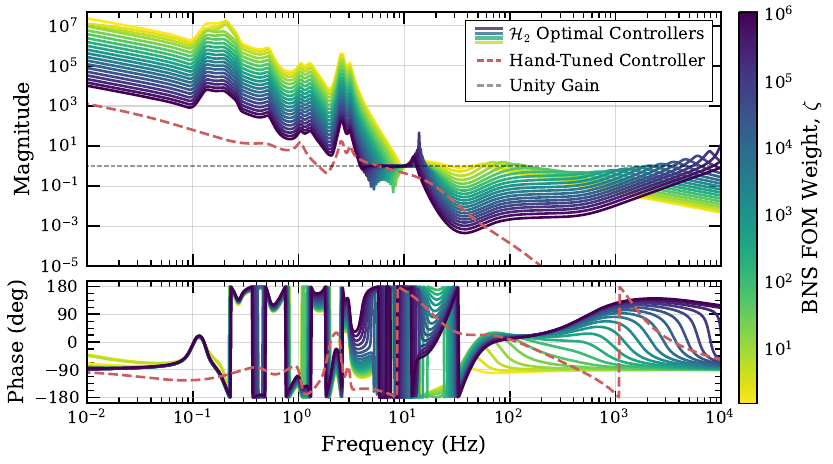}
\caption{The open-loop gain, $G$, of the unscaled plant and controller system shown for different values of $\zeta$. As $\zeta$ gets lower, the optimizer gives less weight to the $\rms_\mathrm{BNS}$. A hand-designed \ac{LIGO} DHARD yaw controller is included for comparison. Each of these \ac{LQG} controllers was optimized using an \ac{LQG} solver with no enforced stability margins. The hand-tuned controller meets the stability requirements necessary for implementation in the interferometer.}
\label{fig:OLplot H2}
\end{figure*}

The open-loop gain of the plant and controller system for the same \ac{LQG} controllers is shown in Fig.~\ref{fig:OLplot H2}. There are several features and similarities demonstrated in both the optimal \ac{LQG} controllers and the \ac{LIGO} hand-tuned controller.

\begin{figure*}
\centering
\includegraphics[width=0.7\linewidth]{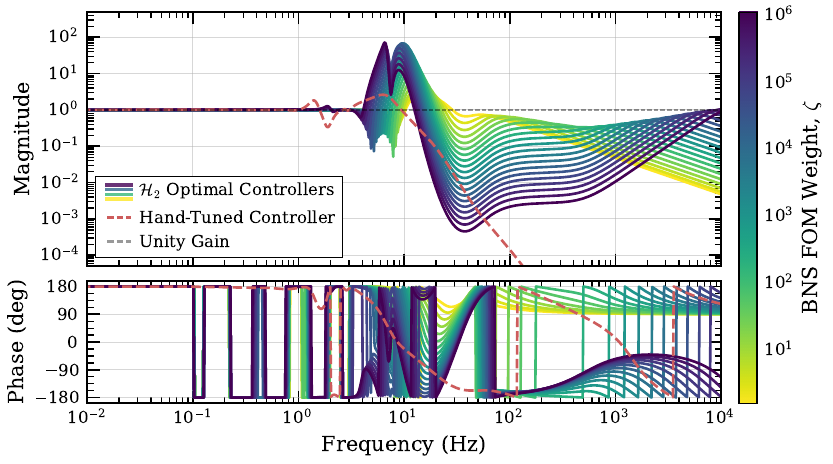}
\caption{The closed-loop gain, $G/(1-G)$, of the plant and controller system shown for different values of the relative \ac{FOM} weighting parameter, $\zeta$. These are the same \ac{LQG} controllers presented in Figs.~\ref{fig:RMSplot} and~\ref{fig:OLplot H2}. The current hand-tuned \ac{LIGO} DHARD yaw controller is included for comparison. The large peaks seen around 10~Hz are gain peaking, which causes low stability margins by inflating the $\mathcal{H}_\infty$ norm of the closed-loop system. This is further discussed in Section~\ref{sec: LQG with arb margins}.}
\label{fig:CLplot H2}
\end{figure*}

The closed-loop gain of $G/(1-G)$, shown in Fig.~\ref{fig:CLplot H2}, is the filter through which the noise $S_{\mathrm{meas}}$, is seen in $S_a$ and $S_h$ as described by Eq.~\ref{eq:CLControlsnoise}. When the peak of $G/(1-G)$ is high, the stability margins are poor, which is visible in Fig.~\ref{fig:CLplot H2}. This suggests that if one can constrain the height, then one can enforce more acceptable margins.

The stability margins are computed for the open-loop gain, $G$, of the \ac{LQG} controllers. The phase margin, which is related to the peak of $G/(1-G)$, is presented as the color of the points in Fig.~\ref{fig:RMSplot H2}. All of the gain and phase stability margins for the \ac{LQG} controllers are unacceptably low, below 10 degrees, where 30~degrees is usually taken to be a lower limit. \ac{LIGO}'s control systems must be robust and remain stable when dealing with drifting operating conditions of the interferometer. If the gain of the optical or mechanical components of the interferometer plant is increased or decreased by a small margin, then the gain of $G$ will also be affected. If the margins are smaller than the change in the gain or phase of the plant, then the system will become unstable.

\subsection{Properties and Limitations Apparent in LQG Controllers}
\label{sec:Properties and Limitations apparent in LQG Controllers}
All of the calculated controllers aim to reduce the noise of the current hand-tuned \ac{LIGO} DHARD yaw controller. Figs.~\ref{fig:RMSplot H2}-\ref{fig:CLplot H2} show the \ac{RMS}, open, and closed-loop transfer functions for the \ac{LQG} controllers and for the hand-tuned \ac{LIGO} controller. The properties exhibited by the \ac{LQG} controllers should be related to the hand-tuned controller, but might also inform future design.

The pieces of information that motivate the hand-tuning of the controllers are generally the \ac{RMS} of $S_p$, the open, and closed-loop transfer function to see the stability margins, as well as the noise spectrum presented in Fig.~\ref{fig:ASCfit}. The hand-tuned controller has an integrator under 1~Hz to suppress the most significant contribution to the total \ac{RMS}: tilt noise. The unity-gain frequency of the open-loop gain is below 10~Hz, and the hand-tuned controller rolls off rapidly above 10~Hz to decrease the impact on the observing range. The roll-off enforces a hand-tuned version of the $F_\mathrm{BNS}$ weighting, $\zeta$, but costs a lot of phase at the unity-gain frequency. The hand-tuned controllers sometimes are designed with multiple unity-gain frequency crossings to target noise peaks where needed, but these are challenging to keep stable, so hand design often just suppresses the lowest-frequency noise.

The \ac{LQG} controllers have a much more aggressive open-loop gain. There is a strong roll-off after 10~Hz to minimize the impact on the \ac{BNS} range, meaning the rate of the roll-off is highly dependent on $\zeta$. When $\zeta$ is very low and thus the $F_\mathrm{BNS}$ does not contribute much to the total \ac{RMS} of $\rms_{\mathrm{syn}}$, the roll-off is almost nonexistent. The limit on when the roll-off becomes non-negligible is when $\zeta \rms_\mathrm{BNS}$ becomes larger than the residual environmental noise $\rms_\mathrm{p}$.
One of the most noticeable characteristics of the \ac{LQG} controllers is that they are aggressive. The closed-loop gain, shown in Fig.~\ref{fig:CLplot H2}, shows that the unchecked gain peaking can get to about 70 times its DC gain, which corresponds to an unacceptable phase margin of around a few degrees.

\section{Solution With Custom Stability Margins}
\label{sec: LQG with arb margins}
While the goal of many of \ac{LIGO}'s control loops is to minimize the $\mathcal{H}_2$ norm of the closed-loop system, optimally, for bilinear noise, the controllers must also be robust and stable. When hand-designing control loops, the measures of this robustness are the gain and phase margins. The gain margin is the smallest gain shift that makes the system unstable, and the phase margin is the smallest amount of phase lag that causes the system to become unstable.

The pure \ac{LQG} solutions for the system, presented in Section~\ref{subsec:LQG Results Simple}, show unacceptable stability margins and gain peaking. The best controller for many of \ac{LIGO}'s applications would be an \ac{LQG} controller if there were a way to guarantee its stability margins. Unfortunately, to quote the entire abstract of J. Doyle's paper on guaranteed margins for \ac{LQG} controllers, ``There are none''~\cite{DoyleITAC78GuaranteedMargins}.

We must instead employ a more complicated strategy. The phase margin for the system is directly linked to the peak of the closed-loop gain. This peak can be measured using an $\mathcal{H}_\infty$ norm and can be bounded. The following work mixes such a bound within the \ac{LQG} solution and presents robust and noise-optimal controllers for our example alignment (\ac{ASC}) system, and can be used for many other \ac{LIGO}'s other control systems.

\subsection{$\mathcal{H}_\infty$ Optimization Problems}
\label{subsec: h_inf simple explination}

Before describing the mixed-sensitivity approach, it is worth first describing the pure $\mathcal{H}_{\infty}$ optimization approach and contrast it to the $\mathcal{H}_2$ minimization performed by \ac{LQG}. In changing an \ac{LQG} problem to utilize an $\mathcal{H}_{\infty}$-optimal solver, the same transfer function $\mathbf{T}_2(f)$ of noise inputs to \ac{FOM} outputs can be used. The optimization objective is then to find the controller that minimizes the maximum value of $\|\mathbf{T}_2(f)\|_2$ over all frequencies $f$ (cf. Section~\ref{subsec:calcHinf}). $\mathcal{H}_{\infty}$ optimization is typically claimed to find controllers more robust to model changes or adversarial inputs; however, it requires its own precise and uniquely defined \acp{FOM} to be applicable or relatable to astrophysics metrics of gravitational-wave detectors. In particular, $\mathcal{H}_{\infty}$ makes no guarantees about minimizing \ac{RMS} noise levels and is thus fundamentally unrelated to SNR and detection range metrics.

In effect, when using noise shaping filters and \ac{FOM} weights, a simple reinterpretation of an $\mathcal{H}_2$ model into  $\mathcal{H}_{\infty}$ neither improves robustness nor helps with the true noise optimization problem. One must instead use a different set of weights and shapes to define a $\mathbf{T}_{\infty}$ transfer function matrix that is dedicated to the $H_{\infty}$ optimization. The following sections show the mixed-sensitivity approach that retains the $\mathbf{T}_2(f)$ of the $\mathcal{H}_2$ optimization while including a dedicated weighting to apply an $\mathcal{H}_{\infty}$ bound. In the following, $\mathbf{T}_{\infty}$ is implemented by utilizing the additional input: $u_{\infty}$ of Fig.~\ref{fig:systemlayout}.

What, then, is the difference between an $\mathcal{H}_{\infty}$ optimization and a bound? Both applications of $\mathcal{H}_\infty$ use an additional single parameter, $\gamma$, to apply the bounded real lemma~\cite{Zhou96RobustOptimal} and produce algebraic Riccati equations that only have an admissible solution when $|\mathbf{T}_{\infty}(f)|^2_2 < \gamma^2$ for all $f$. The $\mathcal{H}_{\infty}$ optimization problem is to find the smallest $\gamma$, and associated controller, such that these equations hold. In finding the minimal admissible $\gamma$, there is no remaining freedom in the controller to also optimize an \ac{LQG} $\mathcal{H}_2$ problem and minimize noise \ac{RMS} due to the uniqueness of the optimal $\mathcal{H}_{\infty}$ controller. When a $\gamma$ above the minimum value is chosen, there is a large class of $\mathcal{H}_{\infty}$-sub-optimal controllers~\cite{GloverS&CL88StatespaceFormulae}, all related through a $\gamma$-bounded linear fraction transformations~\cite{Zhou96RobustOptimal}, that can be considered as the admissible set for an $\mathcal{H}_2$ noise optimization problem. In this way, it is possible to find a solution with a minimal $\mathcal{H}_2$ norm within the set of $\mathcal{H}_{\infty}$-sub-optimal controllers.

\subsection{LQG Optimal Solutions Given an $\mathcal{H}_\infty$ Performance Bound}
\label{sec:results from mix}

As discussed in Section~\ref{subsec: h_inf simple explination}, placing a limit on a suitably defined $\mathcal{H}_\infty$ norm can enforce stability margins of a system. However, $\mathcal{H}_\infty$ solvers do not return controllers with acceptable noise. The ideal controller for most of \ac{LIGO}'s control loops would be the optimal \ac{LQG} filter that has an $\mathcal{H}_\infty$ norm enforcing robustness.
Here, a bound on the phase margin is derived, and results from the alignment control model are shown.
Section~\ref{subsec:mixedH2Hinf} describes the numerical method for calculating the optimal solution~\cite{BernsteinITAC89LQGControl, HaddadP2ICDC89GeneralizedRiccati}.

The phase margin, $\phi_m$, of the system is expressed as the phase when the magnitude crosses the unity-gain frequency, or expressed as an equation,
\begin{align}
    \phi_m = \angle G(f_{\mathrm{u}}) \mathrm{,}
\end{align}
where $f_{\mathrm{u}}$ is the unity-gain frequency. As the unity-gain frequency is approached
\begin{align}
\lim_{f \rightarrow f_{\mathrm{u}}}|G(f)| = 1 \mathrm{,}
\end{align}
meaning, at $f_{\mathrm{u}}$, the closed-loop gain is at or near its maximum. If a bound is added to the closed-loop gain and the phase margin is considered in radians, then
\begin{align}
  \left|\frac{G(f)}{1-G(f)}\right|
  \le \gamma
\end{align}
and
\begin{align}
    \lim_{f \rightarrow f_{\mathrm{u}}} \left|\frac{G(f)}{1-G(f)}\right| \approx \frac{1}{|\phi_m|} \text{,}
\end{align}
yeilding
\begin{align}
                   |\phi_m| &\ge \frac{1}{\gamma}\text{.}
\label{eq:pm<gamma}
\end{align}
By limiting how large $\gamma$ can be through the controller, the phase margin is also limited. For example, $\gamma \approx 1.27$ limit enforces a $\pi/4 = 45$ degree phase margin. 

To implement this, the additional $u_\infty$ input of Fig.~\ref{fig:systemlayout} is employed. By using the transfer function matrix from the input $u_\infty$ to the $y_{_{\mathrm{FOM1}}}$ output to define $\mathbf{T}_{\infty}(f)$, the ideal $\mathcal{H}_{\infty}$-bound can be established in the definitions above.

\begin{figure*}[!htb]
\centering
\includegraphics[width=0.78\linewidth]{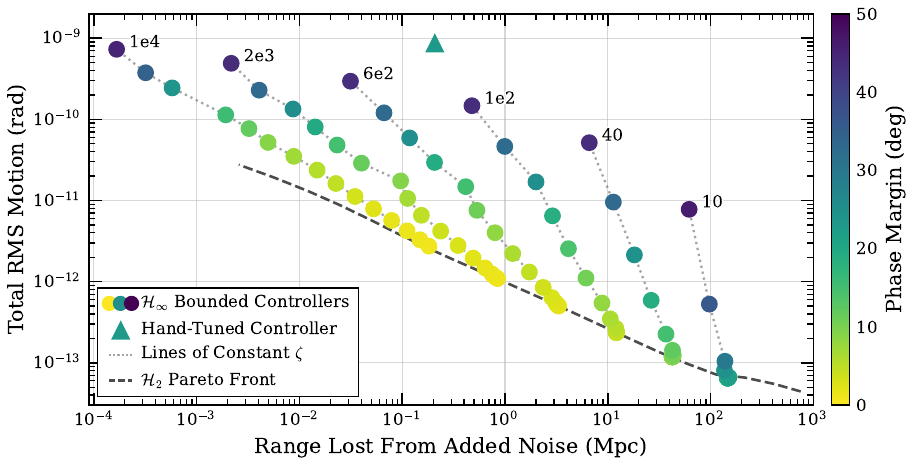}
\caption{The \ac{RMS} plot that shows the \ac{RMS} motion of the optic in radians and the range lost from added noise, the output from both \acp{FOM}. The plot shows the phase margin of all the controllers as a color-coded representation. The phase margin is related to the gain margin monotonically, and their relation as well as their relation to $\gamma$ is presented in Fig.~\ref{fig:gmpm_over_gamma_Plot}. Each branch of the plot, shown by the light dotted line, has the same $\zeta$ value with different $\gamma$ values. The number at the top of each branch is the $\zeta$ value for that branch. The dark dashed line shows the $\mathcal{H}_2$ Pareto front, which is the limit to the controller's noise performance, as it represents the performance of the \ac{LQG} controllers from Fig.~\ref{fig:RMSplot H2}, which are \ac{RMS} optimal.}
\label{fig:RMSplot}
\end{figure*}

\begin{figure*}[!htb]
\centering
\includegraphics[width=0.7\linewidth]{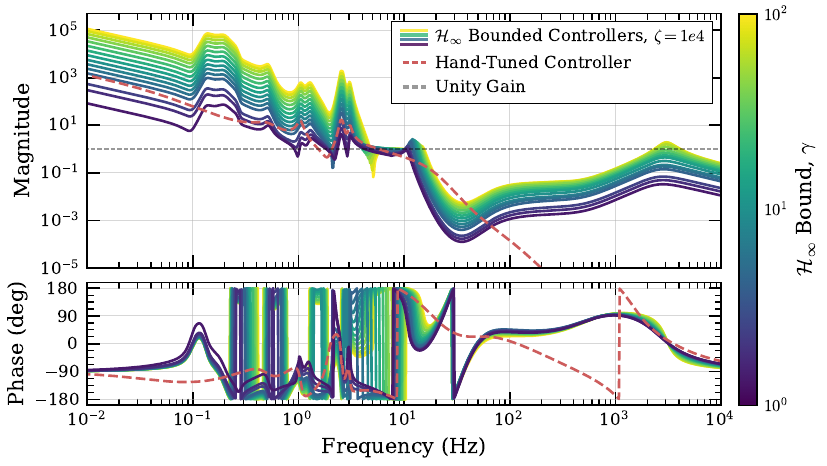}
\caption{The open-loop gain, $G$, of controllers from Fig.~\ref{fig:RMSplot} with a constant $\zeta$ value but with varying $\gamma$ values. As $\gamma$ decreases, the robustness of the controller, as measured by the stability margins, increases. These calculated controllers are compared to the current \ac{LIGO} controller for DHARD yaw.}
\label{fig:OLplot}
\end{figure*}

\begin{figure*}[!htb]
\centering
\includegraphics[width=0.7\linewidth]{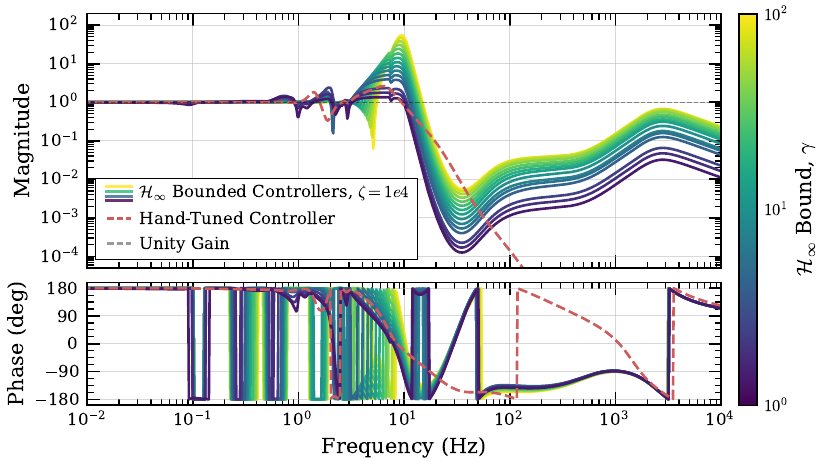}
\caption{The closed-loop gain, $G/(1-G)$, of controllers from Fig.~\ref{fig:RMSplot} with a constant $\zeta$ value but with varying $\gamma$ values. As $\gamma$ decreases, the robustness of the controllers increases. The phase margin of the controller is directly linked to the height of the maximum peak of the closed-loop gain. As $\gamma$ decreases, this peak is suppressed, as shown here, and thus, the stability margins increase. This relation is shown in Fig.~\ref{fig:gmpm_over_gamma_Plot}. These calculated controllers are compared to the closed-loop gain of the current \ac{LIGO} controller for DHARD yaw.}
\label{fig:CLplot}
\end{figure*}

The results, presented in Figs.~\ref{fig:RMSplot},~\ref{fig:OLplot}, and~\ref{fig:CLplot}, show a number of controllers of varying $\zeta$ and $\gamma$ values. Shifting these two parameters allows a group of controllers to be calculated. A single preferred controller can then be post-selected from the group to meet the desired outcome. These plots look similar to those in Section~\ref{sec:LQG Solution Simple}, but include a varying $\gamma$ bound as a parameter.

Interestingly, our configuration of $\mathbf{T}_{\infty}(f)$ to measure the closed-loop gain allows even $\gamma=0$ to be admissible, which occurs when $G=0$, everywhere. In practice, no interesting control loops that influence or optimize the \ac{RMS} can be created as $\gamma \rightarrow 1$. Numerically, our solver becomes increasingly fragile in this limit. Our procedure is to first choose a moderately large $\gamma$ value, and then solve for a controller. If this solve is successful, then it chooses a smaller value and tries again. $\gamma$ is successively reduced towards $\gamma=1$ until it cannot reliably converge to a solution. For this \ac{ASC} example, phase margins of around $45$~degrees are achieved.

The controllers' \ac{RMS} performance, presented in Fig.~\ref{fig:RMSplot}, shows that the current \ac{LIGO} controller for DHARD yaw can be improved in both the \ac{BNS} range as well as the overall \ac{RMS} in the loop while keeping the same phase margin. In the lost range due to the added noise, an order-of-magnitude improvement can be made. The total \ac{RMS} can only be improved by about four times for controllers that would be implemented while maintaining the controller's stability.

Each group of controllers in Fig.~\ref{fig:RMSplot} that is connected by a light dotted line represents controllers with the same $\zeta$, displayed at the top of each branch, for a varying $\gamma$. The phase margin is shown by the color of each point. It is clear that as the phase margin increases, so does the RMS noise. In effect, every choice of phase margin produces a Pareto front.

The open-loop gain of a single branch of these constant $\zeta$ controllers is presented in Fig.~\ref{fig:OLplot} and compared to the current \ac{LIGO} DHARD yaw controller. By plotting only one $\zeta$ value, it is possible to show the effect that $\gamma$ has on the controllers. As $\gamma$ is decreased, the controller becomes more robust, and the overall amplitude decreases, especially at low frequencies, making the controllers less aggressive. This is also true for the other controllers at other $\zeta$ values. However, at around 7~Hz, the response stays almost constant for the varying $\gamma$ values. The fact that the controller is trying to stay at or above unity-gain shows that it is not merely pushing down the gain everywhere but actually trying to maintain the noise performance while improving stability. In comparison to the non-$\mathcal{H}_\infty$ norm limited controllers in Fig.~\ref{fig:OLplot H2}, these are much less aggressive at low frequencies. Between 0.1~Hz and 1~Hz, the gain of the optimal controllers is still sometimes significantly larger than the current \ac{LIGO} controller. In Fig.~\ref{fig:ASCfit}, the environmental noise shaped by the plant is shown to peak around 0.15~Hz. The controller's aggressiveness is trying to suppress the environmental noise as seen at the plant output.

The closed-loop transfer function of $G/(1-G)$ for the system is shown in Fig.~\ref{fig:CLplot}. The $\mathcal{H}_\infty$ norm for each closed-loop system is the maximum of the closed-loop magnitude. Since the $\mathcal{H}_\infty$ norm must always be less than $\gamma$, as $\gamma$ is decreased, the peak of the magnitude of the closed-loop transfer function is forced down. As shown in Eq.~\ref{eq:pm<gamma}, this sets a limit on the phase margin.

\begin{figure*}[!htb]
\centering
\includegraphics[width=0.71\linewidth]{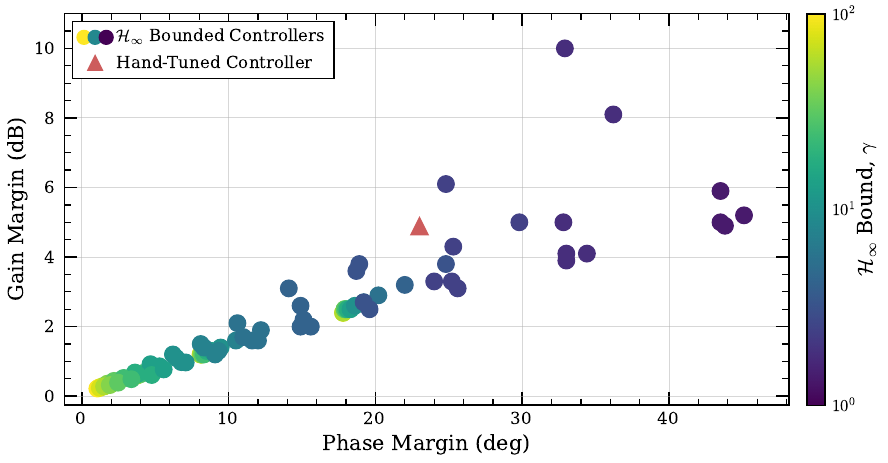}
\caption{The gain margin, phase margin, and $\gamma$ for the \ac{LQG} controllers with $\mathcal{H}_\infty$ performance bounds displayed in Fig.~\ref{fig:RMSplot}. The phase margin, gain margin, and $\gamma$ are monotonic. The stability margins are also shown for the current \ac{LIGO} controller for DHARD yaw, which has no gamma value since it was hand-tuned and not created by an $\mathcal{H}_\infty$ minimization algorithm.}
\label{fig:gmpm_over_gamma_Plot}
\end{figure*}

While the phase margin is directly related to $\gamma$, the gain margin is an important element of the controller's stability. Fig.~\ref{fig:gmpm_over_gamma_Plot} shows that the gain margin and phase margin are monotonic. By putting a bound on $\gamma$, it is possible to also create a controller with both acceptable gain and phase margins. 

\subsection{Properties and Limitations of $\mathcal{H}_\infty$ Norm Limited LQG Controllers}

The first and most important change from the pure \ac{LQG} controllers to the controllers presented in Fig.~\ref{fig:RMSplot}, is the improved phase margin. At the lowest $\gamma$ values, the phase margin can beat the margins of the current hand-tuned controller.

As seen in the closed-loop transfer function, the $\mathcal{H}_\infty$ norm is being suppressed as the $\gamma$ value decreases, enforcing an upper limit. The limiting of the $\mathcal{H}_\infty$ norm is not seen in the plot of the closed-loop gain of the pure \ac{LQG} controllers in Fig.~\ref{fig:CLplot H2}, since there is no $\mathcal{H}_\infty$ boundary enforced. The enforcement of the limit directly leads to more robust controllers as measured by the gain margin and phase margin.

While the \ac{LQG} results in Section~\ref{subsec:LQG Results Simple} show controllers that do not show all of the same features of the current hand-tuned \ac{LIGO} controller, the $\mathcal{H}_\infty$ bounded controllers show many of the same features. In the open-loop transfer function (Fig.~\ref{fig:OLplot}), the roll-off of the hand-tuned and $\mathcal{H}_\infty$ bounded controllers is very similar in frequency and magnitude. This similarity appears even though $\mathcal{H}_\infty$ bounded controllers' \ac{RMS} is calculated from $N_{\mathrm{a}}$ rather than $u_{\mathrm{c}}$, as is often used in \ac{LIGO}'s controller optimization. This similarity may suggest that this distinction is not as important. At very low frequencies, the $\mathcal{H}_\infty$ bounded controllers are actually slightly less aggressive than the hand-tuned controller. As $\gamma$ decreases, the $\mathcal{H}_\infty$ bounded controllers get less aggressive.

A notable difference between the current hand-tuned \ac{LIGO} controller and the calculated $\mathcal{H}_\infty$ bounded controllers is their behavior around unity gain ($G=1$). The hand-tuned controller causes many unity-gain crossings in the open-loop gain as the gain peaks over the line to target noise peaks. Only the $\mathcal{H}_\infty$ bounded controllers with the lowest $\gamma$ values exhibit this behavior.

It is important to note that the solver employed does not exactly implement the gain-peaking constraints that enforce phase margin. Our implementation is slightly more restrictive. Current numerical solution methods for the mixed-sensitivity problem cannot fully independently define $\mathbf{T}_2$ and $\mathbf{T}_{\infty}$. Therefore, our method also has to include the $y_{_{\mathrm{FOM2}}}$ output, which instead implements the constraint
\begin{align}
  \left|\frac{G(f)}{1-G(f)}\right|^2
  &\le \frac{\gamma^2}{|F_{\mathrm{flat}}(f)|^2 + \zeta^2 |F_{\mathrm{BNS}}(f)|^2} \mathrm{.}
    \label{eq:gamma-constraint-true}
\end{align}
This could be corrected using a compensating weighting filter on the $u_{\infty}$ input, but this increases the number of system states, requires more computation, and requires a more advanced set of numerical equations. The existing results are satisfactory, though slightly more constrained than necessary to enforce a stability requirement.

\section{Riccati Equation Approach to Mixed-Sensitivity Controls}
\label{sec:theory}

This paper, so far, presents results and explores the \ac{LQG}, $\mathcal{H}_\infty$, and mixed \ac{LQG}/$\mathcal{H}_\infty$ controls problem from the view of classical controls. To numerically solve these problems, the tools and methods of modern controls must be employed. This section explains how the above results were obtained in a way that is useful to a reader trying to replicate or further develop the techniques. This section is not essential to understand the results given above. Notably, additional transformations on the system in Fig.~\ref{fig:systemlayout} are required to meet the structure and conditions of an \ac{LQG} problem.

We start with the definitions and rules for establishing a well-posed \ac{LQG} problem, relating them to the physical requirements and necessary arrangements on the state diagrams. Additional considerations are required to enforce and solve the $\mathcal{H}_{\infty}$ bounds. The system is then extended to the mixed-sensitivity problem.

\subsection{State-Space System Description}
\label{sec:State Space Systems}
\begin{figure}[!h]
\centering
\includegraphics[scale=0.85]{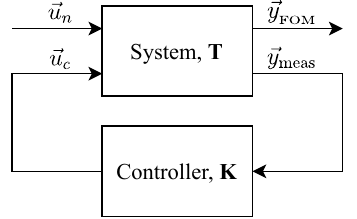}
\caption{A block diagram depicting the connection of the controller to the plant. Including the controller, The transfer function from $\vec{u}_i$, restricted to only the $\vec{u}_{n}$, to $\vec{y}_{_\text{FOM}}$ is referred as $\mathbf{T}_2$ in previous sections. The transfer function from $\vec{u}_i$, restricted to only the $u_{\infty}$, to $\vec{y}_{_\text{FOM}}$ is referred as $\mathbf{T}_{\infty}$.}
\label{fig:controllerdiagram}
\end{figure}

To define an \ac{LQG} problem, the ``System Model'' component of Fig.~\ref{fig:systemlayout} must be assembled into a monolithic state-space. All of the component subsystems are individually modeled or fit from data and converted to state-space form. They are then connected according to the diagram using composition rules, and the entire system, excluding the controller, is put into the form:
\begin{align}
\dot{\vec{x}}  &= \mathbf{A}\vec{x}+{\mathbf{B}_1}{\vec{u}}_i+{\mathbf{B}_2}{\vec{u}}_{\mathrm{c}} \label{eq:SSxdot} \\
{\vec{y}}_{_\mathrm{FOM}}  &= {\mathbf{C}_1}\vec{x}+{\mathbf{D}_{12}}\vec{u}_{\mathrm{c}} \label{eq:SSyfom} \\
  {\vec{y}}_{\mathrm{meas}}  &= \mathbf{C}_2\vec{x}+\mathbf{D}_{21}{\vec{u}}_i+{\mathbf{D}_{22}}\vec{u}_{\mathrm{c}} \label{eq:SSymeas}
\end{align}
where $\mathbf{A}$ is the $n \times n$ state matrix, $\mathbf{B}_j$ is the $n \times m_j$ input matrix, $\mathbf{C}_i$ is the $p_i \times n$ sensing matrix, $\mathbf{D}_{ij}$ is the $p_i \times m_j$ feed-through matrix, $\vec{x}$ is the state vector, ${\vec{u}}_{\mathrm{c}}$ is the control input, ${\vec{u}}_i$ is the vector of inputs for simulating noise in your system, ${\vec{y}}_{_\mathrm{FOM}}$ is the vector of \ac{FOM} outputs, and ${\vec{y}}_{\mathrm{meas}}$ is the measured output. Using $s=2\pi j f$, the system transfer function matrix, $\mathbf{T}$, is given by

% \begin{widetext}
\begin{align}
  \mathbf{T}(f)&=\left[
    \begin{array}{c|cc}
      \mathbf{A} & \mathbf{B}_1 & \mathbf{B}_2 \\
      \hline 
      \mathbf{C}_1 & 0 & \mathbf{D}_{12} \\
      \mathbf{C}_2 & \mathbf{D}_{21} &  \mathbf{D}_{22} \\
    \end{array}\right]\\
  &\equiv
  \begin{bmatrix}
    \mathbf{C}_1 \\ \mathbf{C}_2
  \end{bmatrix}
  \left(s - \mathbf{A}\right)^{-1}
  \begin{bmatrix}
    \mathbf{B}_1 & \mathbf{B}_2
  \end{bmatrix}
  +
  \begin{bmatrix}
    0 & \mathbf{D}_{12}
                   \\
    \mathbf{D}_{21} & \mathbf{D}_{22}
  \end{bmatrix}\text{.}
\label{eq:TFM}
\end{align}
% \end{widetext}
It is important to note that while the system described by $\mathbf{T}$ is often called the ``plant'' in controls literature, in this work it will continue to be called the system model so as to differentiate it from the plant in Fig.~\ref{fig:systemlayout}. Given a $\mathbf{T}$ in the proper form, it is possible to calculate a controller, \controller, that minimizes the $\mathcal{H}_2$ norm of the combined system shown in Fig.~\ref{fig:controllerdiagram}~\cite{DoyleITAC89StatespaceSolutions, ionescu1992l2}.

In the following, the inputs and outputs of the $\mathbf{T}_2$ and $\mathbf{T}_{\infty}$ components of this system will need to be distinguished. Those can be distinguished by segmenting the noise inputs into the white noise and infinity contributions, using
\begin{align}
\mathbf{B}_1 &=
  \begin{bmatrix}
    \mathbf{B}_{1n} & \mathbf{B}_{1\infty}
  \end{bmatrix}\text{,}\\
    \mathbf{D}_{21} &=
                   \begin{bmatrix}
                     \mathbf{D}_{21w} & \mathbf{D}_{21\infty}
                   \end{bmatrix}\text{,}
\end{align}
and
\begin{align}
  \vec{u}_i &=
  \begin{bmatrix}
    \vec{u}_n \\
    u_{\infty}
  \end{bmatrix}\text{.}
\end{align}

\subsection{Calculating The $\mathcal{H}_2$ Norm}
\label{subsec:calcH2norm}
With the state-space representation of the system, we can use it to calculate the total RMS noise given any controller. Along with evaluating the transfer function equations of Eq.~\ref{eq:TFM} to construct the frequency-resolved noise budget, we can use matrix mathematics from numerical Lyapunov solvers to efficiently calculate the total \ac{RMS} noise without numerically evaluating integrals\footnote{this is well known in the control systems community, but less well known in the GW interferometer control commissioning community.}. This technique is used to construct the plots of Figs.~\ref{fig:RMSplot H2} and~\ref{fig:RMSplot}.

The $\mathcal{H}_2$ norm of the system is a measure of the \ac{RMS} at ${\vec{y}}_{_\mathrm{FOM}}$ output from the white noise input ${\vec{u}}_n$. For a continuous system with a transfer function, $\mathbf{T}_2(s)$, the $\mathcal{H}_2$ norm is given by
\begin{equation}
\left\| \mathcal{H}_2 \right\| = \sqrt{\frac{1}{2\pi}\int_{-\infty}^{\infty}{\mathrm{trace[}\mathbf{T}_2\left(j\omega\right)}^\trans \mathbf{T}_2\left(j\omega\right)]d\omega} \mathrm{.}
\label{eq:h2norm}
\end{equation}
The $\mathcal{H}_2$ norm is calculated by solving the Lyapunov equation given by
\begin{equation}
 \mathbf{A}^\trans L_o+L_o\mathbf{A}+\mathbf{C}^\trans\mathbf{C}=\mathbf{A}L_{\mathrm{c}}+L_{\mathrm{c}}\mathbf{A}^\trans+\mathbf{B}\mathbf{B}^\trans=0 \mathrm{,}
\label{eq:lyap}
\end{equation}
where $L_o$ is the observability Gramian of ($\mathbf{C}$, $\mathbf{A}$), $L_{\mathrm{c}}$ is the controllability Gramian of ($\mathbf{A}$, $\mathbf{B}$). Both are determined by numerically solving the Lyapunov equations. In this expression, the matrices $\mathbf{A}$, $\mathbf{B}$, $\mathbf{C}$ are the matrices of the $\mathbf{T}_2$ system whose $\mathcal{H}_2$ norm is being measured (usually with a controller solved and attached). The $\mathcal{H}_2$ norm can then be calculated by
\begin{equation}
\left\| \mathcal{H}_2 \right\| = \sqrt{\mathrm{trace}\left[ \mathbf{B}^\trans L_o \mathbf{B} \right]} = \sqrt{\mathrm{trace}\left[ \mathbf{C} L_{\mathrm{c}} \mathbf{C}^\trans \right] } \mathrm{.}
\label{eq:rmslyap}
\end{equation}
Note that these Lyapunov equation methods apply only when $\mathbf{A}$ is stable, but that condition is guaranteed for the closed-loop systems returned by \ac{LQG}-based solvers. Allowing $\mathbf{A}$ to be unstable changes the interpretation to an $\mathcal{L}_2$  norm calculation, and can be solved using the SLICOT function \texttt{AB13BD}~\cite{varga1ACC92Computing2Norms}. We have noticed that standard Lyapunov equation solvers sometimes fail to produce accurate noise estimates due to the dynamical range of the transfer functions involved. More advanced Lyapunov solvers that compute the ``square root'' of the noise covariance matrices using Hammarling's method and return Cholesky decompositions are more numerically reliable~\cite{AntoulasAoLDS056Sylvester, HAMMARLINGIJoNA82NumericalSolution}.

The use of Lyapunov equations to compute the $\mathcal{H}_2$ norm is foundational to \ac{LQG} theory and can be used to derive the following Riccati equations for optimal control~\cite{AthansITAC67DirectDerivation}. While this is widely known in the controls community and textbooks, we include this here for a physics audience.

\subsection{Optimal LQG Controllers}
\label{subsec:optController}

\newcommand{\KLQG}{\ensuremath{\overline{K}}}
\newcommand{\XLQG}{\ensuremath{\overline{X}}}
\newcommand{\YLQG}{\ensuremath{\overline{Y}}}

Here, we outline some of the major points of the \ac{LQG} numerical solution process, as it will inform the mixed-sensitivity method to follow.

The optimal \ac{LQG} controller, $\mathbf{K}_{\mathrm{LQG}}$, is found by solving two continuous algebraic Riccati equations,
\begin{equation}
\begin{split}
\mathbf{A}^\trans \XLQG+\XLQG\mathbf{A}-\left(\XLQG\mathbf{B}_2+\mathbf{C}_1^\trans \mathbf{D}_{12}\right)\left(\mathbf{D}_{12}^\trans \mathbf{D}_{12} \right)^{-1}\quad\quad\\
 \left( \mathbf{B}_2^\trans \XLQG+\mathbf{D}_{12}^\trans \mathbf{C}_1 \right)+\mathbf{C}_1^\trans \mathbf{C}_1=0 \mathrm{,}
\end{split}
\end{equation}
and
\begin{equation}
\begin{split}
\mathbf{A}\YLQG+\YLQG\mathbf{A}^\trans-\left( \YLQG \mathbf{C}_2^\trans+\mathbf{B}_1 \mathbf{D}_{21}^\trans \right) \left( \mathbf{D}_{21} \mathbf{D}_{21}^\trans \right)^{-1}\quad\quad\\
\left( \mathbf{C}_2 \YLQG+\mathbf{D}_{21} \mathbf{B}_1^\trans \right) + \mathbf{B}_1 \mathbf{B}_1^\trans=0 \mathrm{.}
\end{split}
\end{equation}

Each of these Riccati equations corresponds to the Lyapunov equations of the prior section. The $\YLQG$ is the observability  Gramian for the optimally-controlled full-state feedback system, where the estimator/controller is unknown but assumed to be a Kalman filter. It is also a positive definite symmetric matrix. As the sensing noise becomes smaller, the Kalman gain increases and $\YLQG$ becomes increasingly positive semidefinite, eventually unable to be resolved numerically. Analogously, $\XLQG$ is a positive (semi)definite matrix and the controllability Gramian of the closed-loop system.

The stabilizing solutions, defined below, $\XLQG \geq 0$ and $\YLQG\geq 0$, are then used to calculate the Kalman gain, $K_s$, and the regulator feedback, $F_s$ using:
\begin{equation}
F_s := -\left( \mathbf{D}_{12}^\trans \mathbf{D}_{12}\right)^{-1} \left( \mathbf{B}_2^\trans \XLQG+ \mathbf{D}_{12}^\trans \mathbf{C}_1 \right)
\end{equation}
and
\begin{equation}
K_s :=  -\left( \mathbf{D}_{21} \mathbf{D}_{21}^\trans \right)^{-1} \left( \mathbf{C}_2 \YLQG+\mathbf{D}_{21}\mathbf{B}_1^\trans\right) \mathrm{.}
\end{equation}
The strictly proper controller that solves the $\mathcal{H}_2$-control problem is given by
\begin{equation}
\KLQG =\left[\begin{array}{c|c} 
	{\mathbf{A}}+{\mathbf{B_2}}F_s+K_s^\trans{\mathbf{C_2}} & K_s^\trans \\ 
	\hline 
	-F_s & 0 
\end{array}\right] \mathrm{.}
\end{equation}
Note that the above equations do not use $\mathbf{D}_{22}$, and assume it to be zero. One can remove or add the $D_{22}$ using a feedforward term on $\mathbf{L}$, which can be reinterpreted (and numerically implemented) as a feedback term on $\KLQG$. This amounts to a linear canonical transformation to find the final optimal \ac{LQG} controller $\mathbf{K}_{\mathrm{LQG}}$ of the as-given system:
\begin{align}
  \mathbf{K}_{\mathrm{LQG}} &= \KLQG(\mathbf{I}_{p_2}+\mathbf{D}_{22}\KLQG)^{-1} \mathrm{,}
  \label{eq: nonzero D22 correction}
\end{align}
where $\mathbf{I}$ is the identity matrix of dimension $p_2$.

The existence and numerical solvability of these equations depend on many results in control theory and mathematical analysis. We emphasize two that are relevant to discussing our mixed-sensitivity approach to follow.

The first is the separability of the solutions. The controller $\KLQG$ is computed from two independent sub-problems. The two independent problems are the linear-quadratic observer (a Kalman filter) and the linear-quadratic regulator. Each problem has stability guarantees for the limiting case of state feedback or state estimation, yet these combine to form a single control system that is itself stable and that stabilizes the system under feedback. This separability is a major reason that the solution is non-iterative and can be proven to be globally optimal.

The second key result we emphasize is the development of deflating subspace methods to solve algebraic Riccati equations~\cite{laubITAC79SchurMethod, laubTRE91InvariantSubspace}. In short, these methods transform either of the above equations into a single matrix representing a particular system of equations. The matrix is transformed into a Schur form, upper triangular with its eigenvalues along the diagonal. During the conversion to Schur form, a modified QR or QZ algorithm reorders the stable eigenvalues first along the diagonal. Once in this form, the system of equations can be solved using a column of the orthogonal transformation matrix, the deflating subspace. In the original formulation of Laub, the matrix is ``Hamiltonian'', a form that ensures that half of the eigenvalues are stable. That form can be generalized~\cite{VanDoorenSJSaSC81GeneralizedEigenvalue} to solve directly for the feedback control as well. The choice of eigenvalues represents the poles of the closed-loop system, ensuring stability under optimal control. Under the conditions of the following section, the Gramians \XLQG and \YLQG are ensured to exist and to be positive definite, which in turn ensures that the controller, as an independent system, is itself stable.

We mention these two as they are essential for the solution of the mixed-sensitivity problem and for further improving numerical robustness for high dynamic range gravitational wave system models. For example, the \texttt{ordqz} algorithm that is essential to the Scipy implementation of the deflating subspace method, from which we base our implementations, greatly benefits from matrix rescaling~\cite{James14MatrixBalancing, ParlettNM69BalancingMatrix}. This has led us towards some custom (re-)implementations of these solvers and a desire to test against quad-precision linear algebra implementations, where the physical dynamic range is a substantially smaller fraction of the floating-point epsilon. We leave this for future work and note that several libraries exist in the Julia language, which may make such a future effort possible. In our testing, we have not found that iterative refinement improves upon the reliability of a deflating subspace solver~\cite{Susca2IICAQTRA20IterativeRefinement}.

\subsection{Mathematical Assumptions and Their Physical Interpretation}
\label{subsec:assumptions}

In order to calculate the optimal controllers for the system given in Eqs.~\ref{eq:SSxdot}-\ref{eq:TFM}, the following assumptions must be met to ensure that the problem is well-posed. These are presented mathematically here, followed by the physical requirements and manipulations of our system to generically enforce them.

\begin{enumerate}
  \item\label{itm:whitenoise} The inputs, ${\vec{u}}_n$, must be unit white noise.
  \item\label{itm:stabilizable} ($\mathbf{A}$, $\mathbf{B}_2$) is stabilizable and ($\mathbf{C}_2$, $\mathbf{A}$) is detectable. This assumption is necessary for an internally stable closed-loop system. More restrictive assumptions are given in~\cite{DoyleITAC89StatespaceSolutions}.
  \item\label{itm:Dfullcolumnrank} $\mathbf{D}_{12}$ and $\mathbf{D}_{21}$ both have full column rank.
  \item \label{itm:nonsingularD12}$\mathbf{D}_{12}^\trans \mathbf{D}_{12}^{}$ is invertible and numerically non-singular.
  \item \label{itm:nonsingularD21}$\mathbf{D}_{21}^{} \mathbf{D}_{21}^\trans$ is invertible and numerically non-singular.
  \item \label{itm:D11=0}$\mathbf{D}_{11} = 0$. Note that $\mathbf{D}_{22}$ is often also assumed to be $0$ for use separable solutions, but does not have to be. If $\mathbf{D}_{22} \neq 0$, the optimal controller can be computed as though it is zero, and then transformed as shown in Eq.~\ref{eq: nonzero D22 correction}.
  \item\label{itm:rankB2C1D12} rank$\begin{bmatrix} \mathbf{A}-j \omega \mathbf{I}_n & \mathbf{B}_2 \\ \mathbf{C}_1 & \mathbf{D}_{12} \end{bmatrix} =n+m_2\mathrm{, } \quad \forall \omega \in \mathbb{R}$, 
    where $n$ is the order of $\mathbf{A}$, $m_2$ is dimension of ${\vec{u}}_{\mathrm{c}}$, and $\mathbf{I}$ is the identity matrix also of order $n$.
  \item\label{itm:rankB1C2D21} rank$\begin{bmatrix} \mathbf{A}-j \omega \mathbf{I}_n & \mathbf{B}_1 \\ \mathbf{C}_2 & \mathbf{D}_{21} \end{bmatrix} =n+p_2\mathrm{, } \quad \forall \omega \in \mathbb{R}$, where $p_2$ is the dimension of ${\vec{y}}_{\mathrm{meas}}$.
\end{enumerate}

These formal rules are related to the physical description through the following translations:

\begin{description}
\item[\ref{itm:whitenoise}] This is ensured by using noise shaping filters $E$ and $M$ on the two noise inputs, which unwhitens the noise to admit a physical source. Note that at least one of the two noise shapes must have a non-singular $D$ matrix, indicating that the noise becomes asymptotically white at sufficiently high frequency. This is typically the measurement noise $M$. This is required formally to enforce Assumption~\ref{itm:Dfullcolumnrank}, but is physically necessary to make a well-posed optimization tradeoff between low-frequency system noise and measurement noise.
\item[\ref{itm:stabilizable}] This is required for the control and estimation sub-problems of the separation principle to be individually numerically well-posed. Physically, an observable but unstabilizable state contributes to \ac{RMS} but cannot be controlled to suppress its contribution. Thus, it is either a fundamental problem from its real noise contribution or should simply be ignored. If ignored, model reduction should be used. Because of well-known issues in choosing tolerances, structural algorithms to produce minimal realizations~\cite{VargaK90ComputationIrreducible} are not preferred. Instead, Grammian-based truncation algorithms both reduce systems and improve other numerical properties. We employ singular perturbation approximation-based model reduction ~\cite{VargaIPV93NewSquareroot} using \texttt{AB09ND} from SLICOT~\cite{BennerAaCCSaCV199SLICOTSubroutine}.

  \item[\ref{itm:Dfullcolumnrank}-\ref{itm:nonsingularD21}] These requirements are necessary to make a well-posed minimization problem where the control system couples white noise into its states through $D_{21}$ and then that noise is coupled to the output through $D_{12}$. Without those requirements, there is excessive freedom to choose how the noise is rolled off. Along with requirements on the $E$ and $M$ shaping filters, this also requires that the ``plant'' component has a non-singular D matrix, and thus cannot be strictly proper. Additionally, at least one of the \ac{FOM} filters must have a non-singular D matrix. Accommodating more general strictly proper plant systems is non-trivial but generically possible with the augmentation methods of the following section.
  \item[\ref{itm:D11=0}] The $\mathbf{D}_{11}$ condition is necessary to prevent an infinite noise power from contaminating the problem. The $\mathbf{D}_{22}$ conditions are a mathematical convenience in the problem statement but can be compensated by advanced problem restructuring~\cite{SAFONOVIJC89SimplifyingTheory}.
  \item[\ref{itm:rankB2C1D12}-\ref{itm:rankB1C2D21}] If either of these rank conditions is insufficiently met, then the open-loop gain may have a zero on the imaginary line, and the existence of solutions is not guaranteed, or at least substantially more nuanced. It can often be solved anyway with the numerical reliability of generalized eigenvalue and descriptor-system techniques. See remark 14.1 of~\cite{Zhou96RobustOptimal}.
\end{description}

\subsection{System Augmentation}
\label{subsec:augmentation}

\begin{figure*}[!t]
\centering
\includegraphics[width=0.8\linewidth]{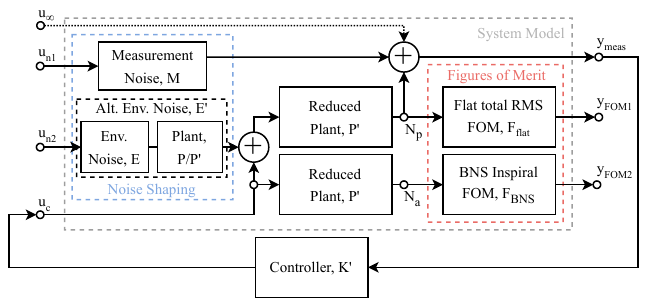}
% \autographicsdrawio{
%   folder=./Graphics/,
%   width=0.8\linewidth,
%   file=ASCSimpSystemOverview_withAug,
% } 
\caption{The layout of the augmented and modified plant system with a controller attached. This alteration of Fig.~\ref{fig:systemlayout} shows the modifications that preserve $\mathbf{T}$ while better ensuring the reliability conditions of Section~\ref{subsec:assumptions}.}
\label{fig:systemlayout_withAugmentation}
\end{figure*}

In order to accommodate the assumptions of Section~\ref{subsec:assumptions}, some augmentations and manipulations are made to make the solver more applicable to general interferometer systems. One example of such augmentations, which has already been discussed, is the inclusion of noise shaping and the \acp{FOM} in Section~\ref{subsec:foms}.
Further manipulations change the system laid out in Fig.~\ref{fig:systemlayout} into a new modified system of Fig.~\ref{fig:systemlayout_withAugmentation}.

To make the new system equivalent to the original, its transfer matrix, $\mathbf{T}(f)$ from the $\vec{u}_n$ inputs to the $\vec{y}_{_\text{FOM}}$ outputs must be preserved. The control inputs and outputs can also be modified during this process, but only in ways that can be stably inverted. For a \ac{SISO} plant and control systems, there is a general method for meeting the assumptions necessary for the solver, as long as the noise shaping filters and \acp{FOM} meet the physical conditions outlined at the end of Section~\ref{subsec:assumptions}.

The following manipulations ensure that Assumption~\ref{itm:nonsingularD12} is met. To do so, much of the plant must be included in the environmental noise block. Most physical plants, such as optical systems or suspensions, have some roll-off at high frequency. This makes them strictly proper systems and implies that the plant subsystem's state-space representation has $\mathbf{D}=0$. By moving most of the plant to the environmental noise block, it can be assured that the remainder has a balanced number of poles and zeros, so that $\mathbf{D}_{12}^{}\neq0$. If the plant is stable, then the only part of the plant that is left in the plant block is the delay or other non-minimum-phase zeros, which can not be moved without changing the physical interpretation. Unstable poles must also remain in the plant section to preserve the problem formulation. The requirement is that the reformulated plant system, $P'$, has $\lim_{f \rightarrow \infty} P'(f) > 0$ while containing the same non-minimum phase zeros and unstable poles as $P$.

Under this redefinition, $P'$ becomes the new plant subsystem, $K'$ is the new calculated controller, and $G'=K'P'$ is the new open-loop gain. Because the $\mathbf{T}_2$ and  $\mathbf{T}_{\infty}$ transfer functions are preserved, the open-loop gain must be the same as the original formulation, thus $G=G'$. After computing $K'$, the controller can be compensated using $K \equiv K'P' / P$ to find the controller of the original formulation. Note that this representation may not be a proper transfer function, and thus may not admit a state-space representation. A simple reason the aforementioned solvers fail is that their outputs are in state-space form, which may not be able to represent optimal controllers, even if the process can properly pose the optimization problem. For \ac{SISO} systems, the final controller may be computed from $K'$ using a zero-pole-gain (ZPK) form or using a descriptor state-space representation.

After transformation, Eq.~\ref{eq:CLControlsnoise} becomes
\begin{align}\label{eq:CLControlsnoise_augmented}
  S_{\mathrm{p}} &= \left |\frac{1}{1-G'}\right |^2  S_{\mathrm{env}}' + \left |\frac{G'}{1-G'}\right |^2  S_{\mathrm{meas}}\text{,}
\end{align}
where
\begin{align}
    S_{\mathrm{env}}'= \left |P/P'\right |^2 S_{\mathrm{env}}
    \mathrm{,}
\end{align}
which is equivalent to the original formulation.

A useful convention for simple plant systems, such as with our \ac{ASC} example, is to make $|P'|=1$ (an inner factor), so that it captures delay and stability effects, without any additional shaping. For this convenient convention, the computed $K' = G$ when the system has no delay, $P'=1$. This convention always has the property $|K'| = |G|$.

If the plant is unstable, all the unstable poles must remain in the plant. To ensure that Assumption~\ref{itm:nonsingularD12} is still met, the stable version of the unstable poles, i.e.~the pole swapped to the left-hand plane, must be added to the plant as zeros. Then those same stabilized poles must be added to the environmental noise system as poles. Since there will be the same number of poles and zeros in $P$, $\mathbf{D}_{22}^{}\neq0$. When executed correctly, the stable poles added to $E'$ will cancel the zeros added to $P'$.

These modifications are essential and relatively simple to implement for \ac{SISO} models. We leave for future work the task of refactoring \ac{MIMO} systems in an analogous manner. We reiterate that the composition of $P'$ uses only the unstable poles and non-minimum phase zeros of $P$. Making $P'$ an \emph{inner coprime} factor of $P$ will satisfy the conditions. We implemented the factorization described above using ZPK representations of \ac{SISO} filters due to the unavailability of inner/outer factorization functions in Python's wrappers of SLICOT (e.g., \texttt{SB08CD}), but anticipate that coprime factorization is effective to extend this formulation for MIMO plants.

In addition to the structural modifications, we also numerically modify the system representation. Given that many individual controls systems in \ac{LIGO} deal with noise over multiple orders of magnitude and models can contain over a hundred states, numerical problems can quickly become a problem for the solvers. To combat this, the system undergoes a matrix balancing stage~\cite{LemonnierSJMA&A06BalancingRegular, WardSJSaSC81BalancingGeneralized}. This better conditions the problems for generalized eigenvalue solvers.
%For the every system presented in this paper, the system was put into Schur form for the mixed \ac{LQG}/$\mathcal{H}_\infty$ solver,
Before solving and after attaching the controller, the system's order was reduced using square-root singular perturbation approximation with preliminary scaling of $\mathbf{A}$, $\mathbf{B}$, and $\mathbf{C}$~\cite{VargaIPV93NewSquareroot, LiuP2ICDC89SingularPerturbation}. We find that these conditioning steps can be essential to solving for complex controllers of gravitational wave interferometer noise models.

\subsection{Calculating the $\mathcal{H}_\infty$ Norm}
\label{subsec:calcHinf}
The $\mathcal{H}_\infty$ norm is a measure of the maximum, peak gain of a closed-loop system over all frequencies.
When used as a reinterpretation of the $\mathbf{T}_2$ transfer function, it can be considered as a measure of a system's robustness against an environment (or adversary) that directs all input noise power to the worst possible frequency. For gravitational-wave detectors, the noise environment is plenty adversarial, but not actively countering our control systems. For specific $\mathbf{T}_{\infty}$ designs, it can be used to compensate specific, known perturbations to $\mathbf{A}$ of the plant system.

Our design of the  $\mathbf{T}_{\infty}$ transfer function, using $u_{\infty}$ to the $\vec{y}_{\mathrm{FOM}}$ outputs, specifically targets the phase margin, just like hand-tuned classical controls designs.
The $\mathcal{H}_\infty$ norm represents the maximum possible gain that is achievable by the system given all possible disturbances.

For a continuous system with a transfer function, $\mathbf{T}_{\infty}(s)$, the $\mathcal{H}_\infty$ norm is most formally defined as
\begin{equation}
\left\| \mathcal{H}_\infty \right\| = \sup_{\omega} \sigma_{\mathrm{max}} \left[\mathbf{T}_{\infty}(j\omega) \right] \mathrm{,}
\label{eq:hinfnorm}
\end{equation}
where $\sup$ is the supremum over $\omega$ and $\sigma_{\mathrm{max}}$ is the maximum singular value of the matrix $\mathbf{T}(j\omega)$.

The means to efficiently find this norm is another highly worthwhile result from mathematical controls called the bounded real lemma. Unlike directly solving for the $\mathcal{H}_2$ norm, the $\mathcal{H}_\infty$ norm can be detected by first assuming a bound, $\gamma$, where $\left\| \mathcal{H}_\infty \right\|<\gamma$. Thus, finding the optimal $\mathcal{H}_\infty$ controller requires a search for the minimum value of $\gamma$. The notion to test this comes from finding the zeros of $0 = \mathbf{T}_{\infty}(f)\mathbf{T}^{\dagger}_{\infty}(f) - \gamma^2\mathbf{1}$. Which is the cross-spectrum matrix of the system (positive semidefinite), less the $\gamma$ bound.
This equation can be produced from a state-space connected to its time-reversed transpose. Finding the zeros of this equation is equivalent to finding the poles of the system $\gamma^{-2}\mathbf{T}_{\infty}(f)\mathbf{T}^{\dagger}_{\infty}(f)$ that has its output fed back to its input. The $A$ matrix of this feedback system is the following Hamiltonian matrix, $\mathbf{H}$, expressed here using the $\mathbf{A}, \mathbf{B}, \mathbf{C}$ of $\mathbf{T}_{\infty}$,
\begin{equation}
  \mathbf{H}:=
  \left[\begin{array}{cc} 
          \mathbf{A} & \gamma^{-2}\mathbf{B}\mathbf{B}^\trans \\ 
          -\mathbf{C}^\trans\mathbf{C} & -\mathbf{A}^\trans
        \end{array}\right] \mathrm{.}
      \label{eq:hamiltonian}
\end{equation}
It follows that, if there are no eigenvalues of this matrix along the imaginary line, then $\left\| \mathcal{H}_\infty \right\|<\gamma$ for the $\mathbf{T}_{\infty}$ system. Thus, it is possible to choose a value for $\gamma$, construct the Hamiltonian matrix, check the spectrum of the solution, then adjust the gamma value and repeat until the lowest gamma value is found within the desired precision. The above algorithm has been highly refined, and it can be directly computed using the SLICOT routine \texttt{AB13DD}~\cite{bruinsmaS&CL90FastAlgorithm}.

We present this result to a physics audience to indicate that $\mathcal{H}_{\infty}$ bounds can be clearly and concisely stated and can be related to a physically meaningful feedback relation. The mixed-sensitivity approach to follow was originally derived by applying Eq.~\ref{eq:hamiltonian} to a noise minimization problem of Eq.~\ref{eq:rmslyap} using Lagrangian constraints.

Finally, the use of singular values in the statement of the norm, Eq.~\ref{eq:hinfnorm}, may not be intuitive. For our single input of $u_{\infty}$, there can be only one singular value of this matrix, which must be the 2-norm of the vector. This leads to the simple $\gamma$-constraint of Eq.~\ref{eq:gamma-constraint-true} without further concern about the complexity of solving for singular values.

\subsection{Optimal $\mathcal{H}_\infty$ Controllers}
\label{subsec:hinf}
Here we review the solution method for $\mathcal{H}_\infty$ optimal control, to relate it to our mixed-sensitivity approach and as background for a physics audience. The primary takeaway is in how $\mathcal{H}_\infty$ utilizes the parameter $\gamma$ as a bound on its norm, which is also utilized in mixed-sensitivity control.

Similar to the $\mathcal{H}_2$ controller, the $\mathcal{H}_\infty$ controller is also found at the minimum of its respective norm. The $\mathcal{H}_\infty$ norm is a measure of the worst-case scenario for the control system as the maximum gain from an input disturbance to the output over all frequencies. In order to calculate an optimal controller, $\mathbf{K}_{\infty\mathrm{(sub)}}$, for the system given in Eqs~\ref{eq:SSxdot}-\ref{eq:SSymeas}, it must follow the assumptions of Section~\ref{subsec:optController}.\footnote{The solution presented here, adapted from~\cite{DoyleITAC89StatespaceSolutions}, is less general than the solution to the \ac{LQG} problem, given in Section~\ref{subsec:optController}. The Glover-Doyle solution is more general, although it goes beyond the needs of most $\mathcal{H}_\infty$ controllers~\cite{gloverdoyle, hong1996derivation}.}%see https://folk.ntnu.no/skoge/prost/proceedings/ecc03/pdfs/190.pdf for a clear description

Unlike solving for the optimal $\mathcal{H}_2$ controller, the $\mathcal{H}_\infty$ controller, like its norm, has one free parameter, $\gamma$, where $\left\| \mathcal{H}_\infty \right\|<\gamma$. Thus, finding the optimal $\mathcal{H}_\infty$ controller requires a search for the minimum value of $\gamma$.

\newcommand{\KINF}{\ensuremath{\overline{K}_{\infty}}}
\newcommand{\XINF}{\ensuremath{\overline{X}_{\infty}}}
\newcommand{\YINF}{\ensuremath{\overline{Y}_{\infty}}}

A sub-optimal $\mathcal{H}_\infty$ controller is found by solving two continuous algebraic Riccati equations,
\begin{equation}
\begin{split}
\mathbf{A}^\trans \XINF+\XINF\mathbf{A}-\XINF \quad\quad\quad\quad\quad\quad\quad\quad\quad\quad\\ \left(\gamma^{-2}\mathbf{B}_1\mathbf{B}_1^\trans-\mathbf{B}_2\mathbf{B}_2^\trans \right)\XINF+\mathbf{C}_1^\trans \mathbf{C}_1=0
\end{split}
\label{eq:hinfRiccatiObs}
\end{equation}
and
\begin{equation}
\begin{split}
\mathbf{A} \YINF+\YINF\mathbf{A}^\trans-\YINF \quad\quad\quad\quad\quad\quad\quad\quad\quad\quad\\ \left(\gamma^{-2}\mathbf{C}_1^\trans\mathbf{C}_1-\mathbf{C}_2^\trans\mathbf{C}_2 \right)\YINF+\mathbf{B}_1 \mathbf{B}_1^\trans=0 \mathrm{,}
\end{split}
\label{eq:hinfRiccatiFB}
\end{equation}

where $\XINF \geq 0$ and $\YINF \geq 0$ must be stabilizing solutions. They are guaranteed to exist as $\gamma \rightarrow \infty$, but will fail to exist at some minimum $\gamma_{\text{min}}$. The resulting admissible controller is
\begin{equation}
\mathbf{K}_{\infty\mathrm{(sub)}}=\left[\begin{array}{c|c} 
	{\mathbf{A}}_{K\infty} & Z_\infty \YINF \mathbf{C}_2^\trans \\ 
	\hline
	-\mathbf{B}_2^\trans \XINF & 0 
\end{array}\right] \mathrm{,}
\label{eq:hinfK}
\end{equation}
where
\begin{equation}
\begin{split}
{\mathbf{A}}_{K\infty}=
	{\mathbf{A}}+(\gamma^{-2}\mathbf{B}_1^{}\mathbf{B}_1^\trans- \mathbf{B}_2^{} \mathbf{B}_2^\trans) \XINF \quad\quad\quad\\ \quad\quad\quad\quad - Z_\infty \YINF \mathbf{C}_2^\trans \mathbf{C}_2
\end{split}
\end{equation}
and $Z_\infty=(\mathbf{I}_n-\gamma^{-2}\YINF^\trans \XINF)^{-1}$. The optimal $\mathcal{H}_\infty$ controller, $\mathbf{K}_\infty$, is the unique controller where $\gamma$ is at its minimum possible value, $\gamma_{\text{min}}$ for these equations to produce a closed-loop-stable system. The failure to produce a solution below this minimum can happen in several ways~\cite{gahinet1ACC92TroubleshootingStatespace}.

One can inspect the form of Eq.~\ref{eq:hinfRiccatiFB} and notice that as $\gamma \rightarrow \infty$, it recovers the equations and solution of the \ac{LQG} problem. More generally, for the set of sub-optimal controllers $\gamma_{\text{opt}} < \gamma < \infty$, this method of $\mathcal{H}_{\infty}$ solution calculates the ``central'' controller that obeys the $\gamma$ bound while minimizing a $\mathbf{T}_2 = \mathbf{T}_{\infty}$ problem. Thus, for mixed-sensitivity problems for which $\mathbf{T}_2$ should be the same as $ \mathbf{T}_{\infty}$, a simple computation of a sub-optimal controller also solves the mixed-sensitivity problem~\cite{DoyleITAC89StatespaceSolutions}.

For the phase-margin constraint and construction of $\mathbf{T}_{\infty}$ presented in this work, the sub-optimal central controllers of the associated $\mathcal{H}_{\infty}$ optimal control problem lack all information about the environmental and sensing noise weights. They are thus not useful for noise minimization. The following shows how to find noise-optimal solutions within the set of sub-optimal controllers.

\section{Optimal LQG Controllers With a Bounded $\mathcal{H}_\infty$ Norm}
\label{subsec:mixedH2Hinf}

The previous section introduced controls concepts for integration into a mixed-sensitivity solver and defined the relevant state-space representation.

Mixed-sensitivity control offers the best of both \ac{LQG} and $\mathcal{H}_{\infty}$ optimization, minimizing noise while ensuring robust performance. Despite its considerable benefits, it is less commonly known or taught, as it is advanced and algebraically cumbersome to describe and solve.

The methodology of this paper shows that it can be incorporated into existing designs in a straightforward manner through the addition of an $u_\infty$ input, such that end-users will only need to input the various components of Fig.~\ref{fig:systemlayout}, and software can do the rest, shielding the user from the problem description. This then requires numerical algorithms that can reliably solve such mixed-sensitivity problems despite the aforementioned numerical challenges, without requiring careful user tuning. This is what we largely, but not entirely, achieved through the following methods.

For context, mixed-sensitivity control was studied shortly after the development of a reliable $\mathcal{H}_{\infty}$- optimal solution using the separation principle. Interestingly, the more direct method of Lagrangian constraints was used to apply the $\mathcal{H}_{\infty}$ bound to the \ac{LQG} noise minimization problem. This was explored in the two initial papers leading into this subject~\cite{BernsteinITAC89LQGControl, HaddadP2ICDC89GeneralizedRiccati}. These two works establish the necessary conditions as a set of coupled Riccati equations to solve and maintain a weaker form of the separation principle. Shortly thereafter, sufficient conditions were determined~\cite{Doyle1ACC89OptimalControl} and the two algebraic approaches were related as duals of each other~\cite{YehITAC92NecessarySufficient}. A notable aspect of these alternate approaches is that they set up the problem using a similar arrangement of inputs and outputs. Namely, there are three sets of inputs and two sets of outputs, or vice versa. One of each is used for observing and control, and the remainder to define the optimization and bound norms.
For the approach of this work to ensure phase margin, the arrangement of inputs and outputs is why the $\mathbf{T}_2$ and $\mathbf{T}_\infty$ matrices are related and lead to the constraint of Eq.~\ref{eq:gamma-constraint-true}.

Three inputs and outputs are needed to define a fully general problem that decouples the $\mathbf{T}_2$ and $\mathbf{T}_\infty$ transfer matrices, but this removes all remaining algebraic separability.
%It can be posed as a convex optimization problem, but we have not been able to produce reliable convex solvers and have not pursued the general mixed-sensitivity problem.
Since the original algebraic formulations, ongoing work has re-framed the mixed-sensitivity problem using \acp{LMI} on a convex domain~\cite{khargonekarITAC91MixedSub, roteaP13ICDC95GeneralizedSub, roteaRCT95GeneralizedH2}. Our first attempts to optimize our system used this approach, through the MATLAB function \texttt{h2hinfsyn}. This function uses an \ac{LMI} approach. For the systems of this paper, it was only successful for $\zeta=0$ problems, with no \ac{BNS} weighting. It was also unacceptably slow. As to why, we can only speculate that the \ac{LMI} formulation, or the specific numerical implementation and internal numerical tolerances, are not currently structured or balanced to handle the large dynamic range of gravitational-wave interferometer models. Alternatively, they may be particularly unsuited when the $\mathbf{T}_2$ weighted optimization problem is defined so differently than the central controllers of our rather dissimilar $\mathbf{T}_{\infty}$ problem. To produce this work, we utilize the earlier algebraic approach of \ac{BH}~\cite{BernsteinITAC89LQGControl, HaddadP2ICDC89GeneralizedRiccati}.

We use the convention of $u_{\inf}$ being on the input side, as that is required for our system layout. This follows the conventions of~\cite{DoyleITAC89StatespaceSolutions} and is dual to~\cite{BernsteinITAC89LQGControl, HaddadP2ICDC89GeneralizedRiccati}. We otherwise use \ac{BH} equations, as they present algebraic solutions for the most general problem. Our formulation uses the following definitions from \ac{BH}, in our notation, to express the algebraic Riccati equations that follow.
\begin{align}
  \mathbf{R}_{1}        &=  \mathbf{B}_{1n} \mathbf{B}_{1n}^\trans \mathrm{,}
  &\\
    \alpha^2\mathbf{\hat{R}}_2 = \mathbf{R}_{2} &=  \mathbf{D}_{21n} \mathbf{D}_{21n}^\trans \mathrm{,}
  &\\
    \mathbf{R}_{12}       &=  \mathbf{B}_{1n}  \mathbf{D}_{21n}^\trans \mathrm{,}
  \\
  \mathbf{V}_{1}       &=  \mathbf{C}_{1}\mathbf{C}_{1}^\trans \mathrm{,}
  &\\
    \mathbf{V}_{2}       &=  \mathbf{D}_{12}\mathbf{D}_{12}^\trans \mathrm{,}
\intertext{and}
    \mathbf{V}_{12}       &=  \mathbf{C}_{1}^\trans \mathbf{D}_{12} \mathrm{.}
\end{align}
These factors are related to the noise inputs and FOM outputs. More relations to map our formulation back to the BH papers are given in Appendix~\ref{app:BHtranslation}. The expressions below require structural similarities between the $D$ matrices of the \ac{RMS} and $\mathcal{H}_{\infty}$ problems. Namely, they have to be proportional so that the $\beta$ factor can be defined as the ratio, $\beta^2 = \mathbf{D}_{21\infty} \mathbf{D}_{21\infty}^\trans / \mathbf{R}_2$. This is easy to ensure for SISO models, but may be more difficult in MIMO extensions.

The factor $\alpha$ is nominally $1$. The equations below are all expressed in terms of $\mathbf{\hat{R}}_2$ rather than $\mathbf{R}_{2}$ or $\mathbf{R}_{2\infty} = \mathbf{D}_{21\infty} \mathbf{D}_{21\infty}^\trans$, leading to the included factors of $\alpha$ and $\beta$. By choosing $\alpha > 1$, we are effectively rescaling the noise matrices in the problem, which is useful in finding solutions.

The set of equations we then must solve is the following three Riccati equations in the positive semidefinite matrices $\mathbf{Q}$,  $\mathbf{Z}$,  $\mathbf{\hat{Q}}$. Included below are the equations along with some convenience definitions to shorten the Riccati equations and highlight their quadratic form. The first of the equations, solving for $\mathbf{Q}$, is independent of the others, indicating the remaining separability of the problem. The equations in $\mathbf{Z}$,  $\mathbf{\hat{Q}}$ are inter-coupled and both rely on the solution to $\mathbf{Q}$. Their simultaneous solution requires an iterative method. The equations are:
\newcommand{\bhA}{\mathbf{A}^\trans}
\newcommand{\bhAt}{\mathbf{A}}
\newcommand{\bhAstZ}{\overline{\mathbf{A}}_{Z}}
\newcommand{\bhAstQh}{\overline{\mathbf{A}}_{\hat{Q}}}
\newcommand{\bhB}{\mathbf{C}_{2}^\trans}
\newcommand{\bhC}{\mathbf{B}_{2}^\trans}
\newcommand{\bhBt}{\mathbf{C}_{2}}
\newcommand{\bhCt}{\mathbf{B}_{2}}
\newcommand{\bhQ}{\mathbf{Q}}
\newcommand{\bhQa}{\mathbf{Q}_a}
\newcommand{\bhQh}{\mathbf{\hat{Q}}}
\newcommand{\bhRf}{\mathbf{R}_{1}}
\newcommand{\bhRhc}{\mathbf{\hat{R}_2}}
\newcommand{\bhVis}{\mathbf{V}_{1\infty}}
\newcommand{\bhVic}{\mathbf{V}_{2\infty}}
\newcommand{\bhVisc}{\mathbf{V}_{12\infty}}
\newcommand{\bhD}{\mathbf{D}_{22}^\trans}
\newcommand{\bhZ}{\mathbf{Z}}
\newcommand{\bhZp}{\mathbf{Z}'}

\begin{widetext}
\begin{align}
0 &= \bhA \bhQ + \bhQ \bhAt + \bhVis - (\bhQ \bhCt + \bhVisc) \bhVic^{-1} (\bhC \bhQ^\trans + \bhVisc^\trans) \mathrm{,}\\
\bhQa &\equiv \bhQ \bhCt + \bhVisc \mathrm{,}\\
  % \nonumber\\
\bhAstZ &\equiv \bhA - \frac{\beta^2}{\alpha^2\gamma^2} \bhQh \bhRf \mathrm{,}\\
0 &= \bhAstZ^\trans \bhZp + \bhZp \bhAstZ + \bhRf - \bhZp\bigg(\frac{\bhB \bhRhc^{-1} \bhBt}{\alpha^2} - \frac{\beta^4 \bhQh \bhRf \bhQh}{\alpha^4 \gamma^4} - \frac{\beta^2 \bhQa \bhVic^{-1} \bhQa^\trans}{\alpha^2 \gamma^2} \bigg) \bhZp \mathrm{,}\\
  \bhZ &\equiv \bhZp / \alpha^2 \mathrm{,}\\
  % \nonumber\\
\bhAstQh &\equiv \bhA - \bhB \bhRhc^{-1} \bhBt \bhZ \mathrm{,}\\
\intertext{and}
0 &= \bhAstQh \bhQh + \bhQh \bhAstQh^\trans + \bhQa \bhVic^{-1}\bhQa^\trans + \bhQh\frac{\beta^2}{\gamma^2}\bhZ \bhB \bhRhc^{-1} \bhBt \bhZ \bhQh \mathrm{.}
\end{align}
\end{widetext}

Using the solutions ($Q$, $Z$, and $\hat{Q}$) to the above equations, the optimal controller is then computed to be
\begin{equation}
{\mathbf{K}_{\mathcal{H}_{2/\infty}}}=\left[\begin{array}{c|c} 
                                              \mathbf{A}_{\mathrm{K}} & \mathbf{B}_{\mathrm{K}} \\ 
	\hline 
                                              \mathbf{C}_{\mathrm{K}}  & 0 
\end{array}\right] \text{,}
\end{equation}
where
\begin{align}
  \mathbf{B}_{\mathrm{K}} = \mathbf{Q}_a \mathbf{V}_{2 \infty}^{-1}\text{,}\\
  \mathbf{C}_{\mathrm{K}} =-\bhRhc^{-1} \bhBt \bhZ\text{,}
  \end{align}
and
\begin{align}
  \mathbf{A}_{\mathrm{K}} = \bhA + \bhB \mathbf{C}_{\mathrm{K}} - \mathbf{B}_{\mathrm{K}} \bhC - \mathbf{B}_{\mathrm{K}} \bhD \mathbf{B}_{\mathrm{K}}\text{.}
\end{align}
The last term on $\mathbf{A}_{\mathrm{K}}$ accounts for the $D_{22} \ne 0$, so no further adjustment is required.

The challenge in solving these equations is in finding the simultaneous solutions. We note some discussion from~\cite{BernsteinITAC89LQGControl}. As long as $\hat{\mathbf{Q}}$ is positive semidefinite, then the solution for $\mathbf{Z}$ will be noise-optimal and stable, but may not obey the given $\mathcal{H}_\infty$ constraint. Likewise, for any positive semidefinite $\mathbf{Z}$, a solution for $\hat{\mathbf{Q}}$ indicates that the control is stable and the $\mathcal{H}_\infty$ bound is met, although it may not be noise optimal. These equations are solved iteratively using a holonomy method, where parameters are scanned from a parameter space where solutions are reliably found to one where they are more challenging to find. The initial condition of the iteration is for $\hat{\mathbf{Q}}=0$.

Instead of performing holonomy on solutions where $\gamma$ is transitioned from $\infty$ to its final value, we instead transition $\alpha$ from $\infty$ to the value 1 while alternately solving the coupled equations. This method is physically motivated. It is equivalent to scaling the sensing noise from being large and dominant, where loops do not need aggressive shaping, to the final form where sensing noise is many orders of magnitude below the environmental noise, while maintaining $\gamma$ at its nominal desired value. 

\subsection{Controllers Stability and Implementation}
\label{subsec:strongly stable controllers}

While the closed-loop system is guaranteed to be stable with either the $\mathcal{H}_2$ or the $\mathcal{H}_\infty$ bounded controller, the controllers alone are not guaranteed to be stable. We have seen that the outputs of the mixed-sensitivity solver produce unstable $\mathbf{K}_{\mathcal{H}_{2/\infty}}$.

In theory, there is nothing concerning about a controller alone being unstable since the overall stability of the system is determined in the closed-loop case. However, it is possible that some may not want to use controllers that are themselves unstable. Especially when gaining lock in one of the interferometer's \acp{DOF}, a stable controller is beneficial. A controller that is independently stable and whose closed loop is stable is called a ``strongly stable'' controller. The equations for strongly stable controllers have been computed in~\cite{kapila1995h2, wang1996stable}. However, it is beyond the goals of this paper to implement.

We have found that in situations where the solved controller is unstable, simply flipping its unstable poles to the stable side still generally results in a closed-loop stable system. This is performed during a reduction stage to lower the order of the controller for implementation in digital control. We validate that the final system model using an order-reduced controller is still stable and that the phase margin of the order-reduced controller is still acceptable after the reduction step. 

\section{Summary and Future Applications}
\label{sec:applicationsAndFutureWork}

In this work, we developed three new concepts to incorporate modern optimal methods into gravitational wave interferometer control.

First, we established a specific figure of merit for optimizing control noise using existing \ac{SNR} or detection-range computations, interpreted as weighted $\mathcal{H}_2$ norms on the control performance, indicating that LQG control is an appropriate framework for optimizing interferometer noise performance.

Second, we compute a Pareto front between a total plant motion and the \ac{SNR} figures of merit by solving for an optimal controller while adjusting a relative weight, $\zeta$, between two noise-metric outputs of our system model.

Third, we resolve the longstanding issue that \ac{LQG} creates arbitrarily low phase margin loops by incorporating mixed-sensitivity optimal control. In our case by imposing a specific $\mathcal{H}_{\infty}$ constraint on the maximum magnitude closed-loop sensitivity function $G(f)/[1-G(f)]$.

Implementing these methods together poses several challenges due to the dynamic range of weighting filters involved, but we were able to reliably form solutions using algebraic methods to solve Riccati equations and an iterative method to jointly solve the coupled Riccati equations that arise in mixed-sensitivity control.

\subsection{Applications}

There are several applications beyond our alignment control example where our method could optimize gravitational-wave detectors.

One of the most apparent applications for this work is to optimize or analyze all classical SISO-designed controllers in \ac{LIGO}. This becomes an obtainable goal since the majority of the systems that contribute significant controls noise are already modeled with their noise contributions budgeted~\cite{O4SensitivityCapote}. The primary barrier to this is un-modeled cross-couplings between different control loops. While achieving SNR-optimized control for the whole interferometer is an ambitious goal, it is possible that using \ac{SISO} optimizations on each individual control loop can approach the \ac{MIMO} optimized controls noise. Future work to adapt this methodology for noise-optimal \ac{MIMO} designs could reveal whether a gap exists between the purely \ac{SISO} design approach and the intrinsically \ac{MIMO} design in \ac{LIGO}'s systems.

One of the most exciting applications of this work is to automatically update controllers in changing noise environments. The noise sources modeled to produce these controllers are not static. They drift over time. Having a robust controller can help the control system adapt to drifting conditions, but the controller is only optimal against a specific noise model. In this case, it is advantageous to have a control solver that continuously updates its models and solutions to respond to both changing noise and a changing plant. This work constructs automated and optimal designs that eliminate the laborious hand-tuning. This opens the possibility of further automating controller design. Commissioning work then moves from designing control loops to maintaining physical and data-driven models of the interferometer, which are useful for more than just control design.

Currently, it is not known how much controls noise can be eliminated from the interferometer as a whole. As shown for our example in Section~\ref{sec: LQG with arb margins}, there are controllers whose noise can be improved. If, however, it is found that the existing controllers are approaching the minimum possible controls noise in the interferometer, it will become imperative to use additional means to improve the overall sensitivity. If the goal remains to reduce controls noise past what our optimal approach can deliver, then non-traditional controls (AI or nonlinear controllers) or hardware upgrades must then be considered. This work enables one to find the limits of control-loop design for SISO systems and better understand how hardware and control loops interact and trade off. With the automated and optimal design we present, more effective hardware and instrumentation requirements can be established more efficiently. Thus, this work could prove useful in parametric designs and upgrades to gravitational-wave detectors, where noise models require auxiliary loop designs, as we now know how to fully saturate performance and treat control-coupled noises as a fundamental noise source.

\section*{Acknowledgements}
The authors gratefully acknowledge the support of the United States National Science Foundation, as well as Elenna Capote, Matthew Evans, Gabriele Vajente, Jeffrey Wack, Terrence Tsang, and the McCuller Group at Caltech for their helpful comments. \ac{LIGO} operates under Cooperative Agreement No. PHY-1764464. Advanced \ac{LIGO} was built under Grant No. PHY-0823459. This paper has been assigned \ac{LIGO} DCC number LIGO-P2300303.

% \section*{References}
% \bibliographystyle{iopart-num.bst}
\nocite{gloverdoyle,hong1996derivation}
\bibliography{LQGbibliography,Zotero_AllControl2}
% \clearpage
\appendix

\section{The Bilinear Noise Metric and Spectral Convolution}
\label{sec:bilinear_noise_spectrum}

This appendix shows how a multiplicative, bilinear coupling affects the spectrum of a signal. Its result is that the spectrum of a convolution is the convolution of a spectrum. Start with two independent noise channels $N_A$ and $N_B$. Each has a spectral density defined as
\begin{align}
  \braket{N_A(f)N_A(-f - f')} &\equiv \frac{1}{2}S_A(f)\delta(f')
\end{align}
and
\begin{align}
    \braket{N_B(f)N_B(-f - f')} \equiv \frac{1}{2}S_B(f)\delta(f')\text{,}
\end{align}
where their independence gives the property $\braket{N_A(f)N_B(f')}=0$.

Now assume that you have a channel $N_x(t)\equiv N_A(t)N_B(t)$. By the convolution theorem, then
\begin{align}
  N_x(f) &= N_A(f) \circledast N_B(f)\\ 
  &\equiv \int_{-\infty}^{\infty} N_A(f') N_B(f - f')  \rd f'\mathrm{.}
\end{align}

The spectrum of this bilinear product can now be derived.
\begin{widetext}
\begin{align}
  \frac{1}{2}S_X(f)\delta(f')
  &\equiv
    \braket{N_X(f)N_X(-f - f')}
    \\
  &=
    \Braket{
    \int_{-\infty}^{\infty} N_A(f'') N_B(f - f'')  \rd f''
    \int_{-\infty}^{\infty} N_A(f''') N_B(-f - f' - f''')  \rd f'''
    }
    \\
    &=
      \int_{-\infty}^{\infty}
      \int_{-\infty}^{\infty}
      \underbrace{\Braket{
      N_A(f'') N_A(f''')}}_{\frac{1}{2}S_A(f'')\delta(f'' + f''')}
      \underbrace{
      \Braket{
      N_B(f - f'') 
      N_B(-f - f' - f''')   
      }}_{
      \frac{1}{2}S_B(f - f'')\delta(f' + f'' + f''')
      }
      \rd f''\rd f'''
  \\
  &=
    \frac{1}{4}
    \delta(f')
    \int_{-\infty}^{\infty}
    S_A(f'')
    S_B(f - f'')
    \rd f''
\end{align}
\end{widetext}
Thus, for this type of bilinear noise,
\begin{align}\label{eq:intermod_deriv}
S_X(f) &= \frac{1}{2}S_A(f) \circledast S_B(f).
\end{align}

This short derivation should now be applied to the bilinear noise problem of Section~\ref{subsec:foms}.
Now we apply it to the bilinear noise term, $N_X = \chi N_{\text{p}}(t) N_{\text{a}}(t)$ seen in Eq.~\ref{eq: h_gw bilinear}.

We first assume that the \ac{RMS} noise power of $S_{\text{p}}$ is concentrated at low frequencies: i.e.
\begin{align}
  S_{\text{p}}(f) &\approx \rms_{\text{p}} \delta(f+\epsilon).
                     \label{eq: concentration}
\end{align}
for some small $\epsilon$.

Applying the convolution from this derivation then gives:
\begin{align}
  S_{X}(f) &= \chi \frac{\rms_{\text{p}}}{2} \left( S_{\text{a}}(f+\epsilon) +S_{\text{a}}(f-\epsilon)  \right)
= \chi {\rms_{\text{p}}} S_{\text{a}}(f)\mathrm{.}
\end{align}

This final result indicates that $|C| = \chi \rms_{\text{p}}$ as used in Section~\ref{subsec:foms}.

\section{Alternate Noise Arrangement}
\label{sec:alternate_FOM}

Along with the model presented in the main text, there are two major alternate approaches to where to place the \acp{FOM} to model bilinear-type noise couplings, and we also suggest a third approach. Each has its own reasons for implementation.

Another alteration present in this system is the apparent bypass of the plant by the environmental noise. In this configuration, the filter that previously represented only the shaping of the environmental noise it now represents the shape of the environmental noise passing through the plant, $|EP|$. Letting this input noise-shaping filter represent the environmental noise passed through the plant allows it to be connected after the plant, since it already has picked up the shape from the plant. This alteration allows the very useful option of fitting the filter to an open-loop spectrum at $y_\text{meas}$ as is shown in Fig.~\ref{fig:ASCfit}.

%Currently employed by \ac{LIGO} in most of its control minimizations is a slight variation of the method used in this paper.

\begin{figure*}[!t]
  \centering
  \includegraphics[width=0.8\linewidth]{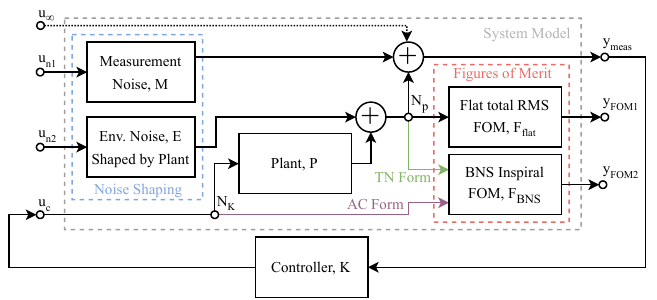}
  % \autographicsdrawio{
  %   folder=./Graphics/,
  %   width=0.8\linewidth,
  %   file=ASCSimpSystemOverview_alt_withNC,
  % }
  \caption{Two alternate system layouts to Fig.~\ref{fig:systemlayout}, the noise-augmented plant system with a feedback controller attached. In this configuration, the environmental noise block is the equivalent of $|EP|$ rather than just $E$ as in all previous system layouts. This change does not alter the system but allows the user to fit an open-loop spectrum at point $y_\text{meas}$ as is shown in Fig.~\ref{fig:ASCfit}. Here, the \acp{FOM} are arranged showing two alternate bilinear noise optimizations. The first, the actuator-centered (AC) form often used by \ac{LIGO}, is where the \ac{BNS} \ac{FOM} is connected to the control signal, $u_{\text{c}}$, and treated as a noise, $N_{\text{K}}$, with the purple connection. The second is the total-noise (TN) form that can be used for non-coupled \acp{DOF} and connects the \ac{BNS} \ac{FOM} to the plant's state noise $N_{\text{p}}$ with a green connection. Only one of these two alternate connection forms can be used at once.
  }
  \label{fig:systemlayout_alt}
\end{figure*}

\subsection{Actuator-Centered form}
\label{subsec:LIGO alternate_FOM}

For the \ac{ASC} bilinear noise model often assumed by \ac{LIGO} controls commissioners, the controls noise arises solely from the $u_{\mathrm{c}}$ feedback contribution to the noise. This type of bilinear noise injection can be modeled by the following coupling.
\begin{align}
  h_{signal}(t) &= h_{\mathrm{gw}}(t) + N_{\mathrm{det}}(t) + \chi N_{\mathrm{p}}(t)u_{\mathrm{c}}(t) \mathrm{.}
              \label{eq: h_gw bilinear_LIGO}
\end{align}
This coupling results when the coupling function is computed from
\begin{align}
|C|^2 = \frac{S_{h\text{, exc}} - S_h}{S_{\mathrm{K}\text{, exc}} - S_\mathrm{K}}\mathrm{,}
\label{eq:A2L coupling2}
\end{align}
where the excited channel in the denominator is the output of the control filter (for alignment, it is related to units of torque on the mirror).

Fig.~\ref{fig:systemlayout_alt} shows how to set up the augmented plant system to implement this alternate form. Note that the injection of the environmental noise is now after the plant, which puts it into the same units as the measurement noise. This makes the injection into DARM appear as:
\begin{align}
  S_{\mathrm{K}} = \left |\frac{G(f)}{1-G(f)}\right |^2  (S_{\mathrm{env}} + S_{\mathrm{meas}})\mathrm{,} %closed loop
  \label{eq:CLControlsnoise_alt}
\end{align}
for this arrangement of noises in the model.

This arrangement is primarily a conceptual change from the model used in the main text. In it, the coupling function takes on different units since it is modeled before entering the plant, and so it is in units of \ac{DARM} length/torque for our alignment control example.

\subsection{Total-Noise Optimization Form}
\label{subsec:non coupled alternate_FOM}
The form and connection point of the \acp{FOM} that is perhaps the most appropriate for quadratically-nonlinear (rather than bi-linear) systems requires using the total noise $S_{\text{p}}$ as discussed in Section~\ref{subsec:foms}. This Setup is detailed in Fig.~\ref{fig:systemlayout_alt} with the connection from $N_{\text{p}}$ to the \ac{BNS} \ac{FOM}. The connection of both \acp{FOM} to $N_{\text{p}}$ ensures that both the actuation point controls noise, as well as environmental noise, are minimized together. This is a more typical arrangement of inputs/outputs and weight filters for LQG problems and does not require two copies of the plant. It corresponds to quadratically nonlinear noise couplings of the form
\begin{align}
  h_{signal}(t) &= h_{\mathrm{gw}}(t) + N_{\mathrm{det}}(t) + \chi N^2_{\mathrm{p}}(t) \mathrm{.}
              \label{eq: h_gw bilinear_LIGO x_N_p}
\end{align}

In terms of numerical implementation, there are several reasons that the \ac{LIGO} preferred forms in Appendix~\ref{subsec:LIGO alternate_FOM} and the main text are less preferable to the total-noise form. One of the major reasons is that in \ac{LIGO}'s form, most of the low-frequency environmental noise shows up in $S_{\text{c}}$ when $|G|$ is large. The dynamic range of the \ac{BNS} \ac{FOM} covers nearly all of the range of a double float. Thus, in computing the noise contribution in this model, there are numerical difficulties multiplying the Environmental noise filter with the \ac{BNS}-weighting filter inside a state-space. The second reason is that bilinear noise is well motivated to be driven by $N_{\mathrm{p}}$, or at least some mixture of $N_{\mathrm{p}}$ and $u_{\mathrm{c}}$. In some instances, \ac{LIGO} has found that further accelerating the roll-off of $G$ does not reduce the noise in \ac{DARM}. The primary noise coupling chosen for this work provides a reason for that result: the bilinear noise can also be driven by residual environmental noise in the plant that is reflected in $N_{\mathrm{p}}$ rather than in $N_{\mathrm{a}}$ or $u_{\mathrm{c}} = N_K$.

\subsection{Measurement-Only Optimization Form}
A final alternate approach would be to use a BNS FOM channel with the noise coupling $S_{\text{c}'} = |G/(1-G)|^2 S_{\text{meas}}$ only, without the environmental noise contribution. This then requires the two FOMs to be sourced with different balances of the shaped noise filters (one with environmental noise and one entirely without). The authors do not know of a means to implement this through the block diagram without two copies of the unknown $K$ control; however, they believe it should be possible to develop a new set of Riccati equations to solve. This might be done by determining the equivalent noise model using a sum of Lyapunov equations, and then solving for the constrained noise minimization problems using the techniques of~\cite{AthansIaC67MatrixMinimum, AthansITAC67DirectDerivation, BernsteinITAC89LQGControl, HaddadP2ICDC89GeneralizedRiccati}. Doing so is likely to further reduce the separability structure expected of \ac{LQG}, and may result in three coupled Riccati equations, rather than the two coupled equations of mixed-sensitivity control. This alternate approach could be primarily useful for numerical implementation reasons, and perhaps would be useful to investigate using a convex optimization approach that can more easily represent and solve constrained optimization problems.

\section{Translation of the \ac{BH} Equations}
\label{app:BHtranslation}

The notation table of~\cite{BernsteinITAC89LQGControl} as well as the notation table and Fig. 1 mapping in~\cite{HaddadP2ICDC89GeneralizedRiccati} can be used to translate the notation of this work into the algebraic equations, through the following mappings:
\begin{align}
  A            &\rightarrow   \mathbf{A}_{}^\trans \mathrm{,}&
  B            &\rightarrow   \mathbf{C}_{2}^\trans \mathrm{,}&
  C            &\rightarrow   \mathbf{B}_{2}^\trans \mathrm{,}&\\
  D            &\rightarrow   \mathbf{D}_{22}^\trans \mathrm{,}&
  {E}_1        &\rightarrow   \mathbf{B}_{1n}^\trans \mathrm{,}&
  {E}_{1\infty}&\rightarrow   \mathbf{B}_{1\infty}^\trans \mathrm{,}&\\
  {E}_{\infty} &\rightarrow  \mathbf{D}_{11\infty}^\trans \mathrm{,}&
  {E}_2        &\rightarrow   \mathbf{D}_{21n}^\trans \mathrm{,}&
  {E}_{2\infty}&\rightarrow   \mathbf{D}_{21\infty}^\trans \mathrm{,}&\\
  {D}_1        &\rightarrow   \mathbf{C}_{1}^\trans \mathrm{,}&
  &\mathrm{and} &
  {D}_2        &\rightarrow   \mathbf{D}_{12}^\trans \mathrm{.}&
\end{align}

Because ${E}_{\infty}$ is structurally zero in the problem setup, the following \ac{BH} parameters are simplified as:
\begin{align}
  \mathbf{M} &= \mathbf{1}\mathrm{,}
  &
    \mathbf{N} &= \mathbf{1}\mathrm{,}
  &
  \\
  \mathbf{V}_{1\infty} &= \mathbf{V}_{1}\mathrm{,}
  &
  \mathbf{V}_{2\infty} &= \mathbf{V}_{2}\mathrm{,}
  &
    \mathbf{V}_{12\infty} &= \mathbf{V}_{12}\mathrm{,}
  \\
  \mathbf{R}_{01\infty} &= \mathbf{0}\mathrm{,}
  &
    \mathbf{R}_{02\infty} &= \mathbf{0}\mathrm{.}
\end{align}
This simplification leads to the reduced expressions
\begin{align}
\mathbf{R}_{1\infty}&= \mathbf{B}_{1\infty} \mathbf{B}_{1\infty}^\trans\mathrm{,}
\\
\mathbf{R}_{2\infty}&=  \mathbf{D}_{21\infty} \mathbf{D}_{21\infty}^\trans\mathrm{,}
\\
\mathbf{R}_{12\infty}&=  \mathbf{B}_{1\infty}  \mathbf{D}_{21\infty}^\trans\mathrm{,}
\end{align}
\begin{align}
  \mathbf{\hat{R}}_2 \equiv \mathbf{R}_2/\alpha^2 = \mathbf{R}_{2\infty} / \beta^2\mathrm{,}
\end{align}
where
\begin{align}
    \alpha \equiv 1 \mathrm{.}
\end{align}

We also assume that there is no additional shaping filter on the $u_{\infty}$ input, causing ${E}_{1\infty}$ term from \ac{BH} to be zero. This leads to the additional simplifications:
\begin{align}
  \mathbf{R}_{1\infty} &= \mathbf{0}
\intertext{and}
  \mathbf{R}_{12\infty} &= \mathbf{0}
\end{align}

Note that all of these are then applied to the full form equations of~\cite{HaddadP2ICDC89GeneralizedRiccati} and then simplified into the $\mathbf{Z}$-form description following Section V of~\cite{BernsteinITAC89LQGControl}.

%\bibliography{LQGbibliography,Zotero_AllControl2}

\end{document}